\documentclass[useAMS,usenatbib]{mnras}
\def\apj{ApJ}
\def\apjs{ApJ}
\def\mnras{MNRAS}
\def\aap{A\&A}
\def\apjl{ApJ}

\def\nat{Nature}

\def\apss{Ap\&SS}

\usepackage{fancyhdr}
\usepackage{graphicx,latexsym}
\usepackage{color}
\usepackage{mathtools}
\usepackage[margins]{trackchanges}
\usepackage{multirow}
\usepackage{subcaption}
\usepackage{ctable}
\usepackage{url}
\usepackage{array}
\usepackage{color,soul}
\usepackage[section]{placeins}
\usepackage{bigints}
\usepackage{hyperref}
\usepackage{amsmath,amssymb}

\newcommand{\GG}[1]{}

\addeditor{LdB}
\addeditor{NS}
\addeditor{BB}
\addeditor{MS}

\usepackage{times}

\newcolumntype{M}{>{\centering\arraybackslash}m{\dimexpr.35\linewidth-2\tabcolsep}}
\newcolumntype{N}{>{\centering\arraybackslash}m{\dimexpr.16\linewidth-2\tabcolsep}}
\newcolumntype{Z}{>{\centering\arraybackslash}m{\dimexpr.23\linewidth-2\tabcolsep}}
\newcolumntype{B}{>{\centering\arraybackslash}m{\dimexpr.20\linewidth-2\tabcolsep}}

\title[Cosmic Rates of BH Mergers and PISNe from Chemically Homogeneous Binary Evolution.]{Cosmic Rates of Black Hole Mergers and Pair-Instability Supernovae from Chemically Homogeneous Binary Evolution}
\author[du Buisson et al.]{L. du Buisson$^{1}$\thanks{E-mail:
lise.dubuisson@chch.ox.ac.uk}, P. Marchant$^{2}$, Ph. Podsiadlowski$^{1,3}$, C. Kobayashi$^{4}$, F. B. Abdalla$^{5}$, \newauthor P. Taylor$^{6,7}$, I. Mandel$^{8,9,10,7}$, S.E. de Mink$^{11,12}$, T.J. Moriya$^{13,8}$ and N. Langer$^{3}$\\
$^{1}$Department of Physics, University of Oxford, Denys Wilkinson Building, Keble Road, Oxford, OX1 3RH, UK.\\
$^{2}$Institute of Astronomy, KU Leuven, Celestijnenlaan 200 D, 3001 Leuven, Belgium\\
$^{3}$Argelander-Institut f{\"u}r Astronomie, Universit\"{a}t Bonn, Auf dem H\"{u}gel 71, 53121 Bonn, Germany.\\
$^{4}$Centre for Astrophysics Research, Department of Physics, Astronomy and Mathematics, University of Hertfordshire, Hatfield AL10 9AB, UK.\\
$^{5}$Department of Physics and Astronomy, University College London, Gower Street, London, WC1E 6BT, UK.\\
$^{6}$Research School of Astronomy and Astrophysics, Australian National University, Canberra, ACT 2611, Australia.\\
$^{7}$ARC Centre of Excellence for All Sky Astrophysics in 3 Dimensions (ASTRO 3D), Australia.\\
$^{8}$School of Physics and Astronomy, Monash University, Clayton, Vic. 3800, Australia\\
$^{9}$The ARC Centre of Excellence for Gravitational Wave Discovery -- OzGrav, Australia\\
$^{10}$School of Physics and Astronomy, University of Birmingham, Birmingham B15 2TT, UK\\
$^{11}$Center for Astrophysics, Harvard-Smithsonian, 60 Garden Street, Cambridge, MA 02138, USA.\\
$^{12}$Anton Pannekoek Institute for Astronomy, University of Amsterdam, 1090 GE Amsterdam, The Netherlands.\\
$^{13}$National Astronomical Observatory of Japan, National Institutes of Natural Sciences, 2-21-1 Osawa, Mitaka, Tokyo 181-8588, Japan.
}

\begin{document}

\date{Accepted XXX. Received XXX; in original form XXX}

\pagerange{\pageref{firstpage}--\pageref{lastpage}} \pubyear{2014}

\maketitle

\label{firstpage}

\begin{abstract}
During the first three observing runs of the Advanced gravitational-wave detector network, the LIGO/Virgo collaboration detected several black hole binary (BHBH) mergers. As the population of detected BHBH mergers grows, it will become possible to constrain different channels for their formation. Here we consider the chemically homogeneous evolution (CHE) channel in close binaries, by performing population synthesis simulations that combine realistic binary models with detailed cosmological calculations of the chemical and star-formation history of the Universe. This allows us to constrain population properties, as well as cosmological and aLIGO/aVirgo detection rates of BHBH mergers formed through this pathway. We predict a BHBH merger rate at redshift zero of $5.8 \hspace{1mm} \textrm{Gpc}^{-3} \textrm{yr}^{-1}$ through the CHE channel, to be compared with aLIGO/aVirgo's measured rate of ${53.2}_{-28.2}^{+55.8} \hspace{1mm} \text{Gpc}^{-3}\text{yr}^{-1}$, and find that eventual merger systems have BH masses in the range $17 - 43 \hspace{1mm} \textrm{M}_{\odot}$ below the pair-instability supernova (PISN) gap, and $>124 \hspace{1mm} \textrm{M}_{\odot}$ above the PISN gap. We investigate effects of momentum kicks during black hole formation, and calculate cosmological and magnitude limited PISN rates. We also study the effects of high-redshift deviations in the star formation rate. We find that momentum kicks tend to increase delay times of BHBH systems, and our magnitude limited PISN rate estimates indicate that current deep surveys should be able to detect such events. Lastly, we find that our cosmological merger rate estimates change by at most $\sim 8\%$ for mild deviations of the star formation rate in the early Universe, and by up to $\sim 40\%$ for extreme deviations.
\end{abstract}




\section{Introduction}
\label{sec:intro}

From its first, second and third observing runs (referred to as the O1, O2 and O3 runs, respectively), the LIGO/Virgo collaboration has reported several detections of gravitational-wave signals from merging binary black holes (BHBHs, \citealt{abbott2019, review2019}), and several more candidate detections\footnote{Gravitational Wave Candidate Event Database: \url{https://gracedb.ligo.org}}. Further BHBH merger candidates have  been identified by other groups \citep{iau2019,Nitz:2019}. These BHBH merger detections match the inspiral and merger of BHs covering a range of masses, the smallest progenitor masses belonging to GW170608 at $11.0^{+5.5}_{-1.7} \textrm{M}_\odot$ and $7.6^{+1.4}_{-2.2} \textrm{M}_\odot$ \citep{smallmerge2017}, and the largest progenitor masses belonging to GW170729 at $50.2^{+16.2}_{-10.2} \textrm{M}_\odot$ and $34.0^{+9.1}_{-10.1} \textrm{M}_\odot$ \citep{review2019}. These detections have provided us with the first direct \mbox{evidence} that binary stellar-mass black hole mergers at the present epoch exist. The estimated local merger rate, provided by the aLIGO and Virgo Collaborations after the second observation run, lies in the range ${53.2}_{-28.2}^{+55.8} \hspace{1mm} \text{Gpc}^{-3}\text{yr}^{-1}$, pointing towards the possibility of many BHBH merger detections per year as the detectors reach their full sensitivity \citep{review2019, a2}.



It was already known that binary systems consisting of stellar remnants that merge in a Hubble time existed --- consider e.g. the double neutron star systems such as the Hulse-Taylor binary \citep{p5,p6} and the double pulsar system PSR J0737-3039 \citep{p7}. BHBHs however, have up to very recently eluded us due to their lack of electromagnetic radiation. Therefore, although one can empirically estimate the merger rate of double neutron star systems based on direct observations of binary neutron star populations \citep{p8, p9, p10, p11}, this is not the case with BHBHs due to the lack of observational evidence. Estimates of BHBH merger rates have therefore historically been very uncertain \citep{p12} and reliant on predictions from population synthesis simulations assuming specific stellar and binary evolutionary \mbox{models.}

As black holes constitute the most massive of the stellar remnants, the gravitational-wave radiation from their mergers will be stronger than that emanating from binary neutron star or binary white dwarf mergers. Indeed, several groups had predicted that BHBH mergers could be the dominant source of detections for LIGO \citep{p13, p14, p15, p16, p12}. With the confirmed gravitational-wave detections of BHBH mergers, we now have the first empirical ``observations" of such events. As more and more detections are made and the population of known BHBH mergers grows, it will become possible to constrain the various proposed formation channels for BHBH mergers \citep{p17, p18, p19, p20, Farr, zevin2017}.

There are different suggestions of possible evolutionary pathways to explain the eventual formation of a BHBH system able to merge at the present epoch. Some examples include classical common-envelope evolution, dynamical formation, and chemically homogeneous evolution (CHE). The first of these involves the classical evolution of isolated binary stars where, in order to progress from the initial stage of binary main-sequence stars to the end product of double compact binaries, highly non-conservative mass transfer or at least one common-envelope (CE) phase is required \citep{smarr1976, p25, p26, p27}. CE evolution, however, is poorly understood \citep{p28}, and the BHBH merger rate estimated by including a CE phase in the formation channel has a large uncertainty \citep{p12}. Dynamical formation of BHBHs requires high stellar densities such as those found in globular clusters or galactic nuclei \citep{p29, p30, p32, p33, kulka}. CHE evolution is possible in very close binaries \citep{p34, p35, p36} that have both stars nearly in contact at the start of the hydrogen burning phase. The stars are then tidally locked to the orbit which enforces rapid rotation, induces rotational mixing and keeps the stars chemically homogeneous during most of the core hydrogen burning phase, dramatically changing their further evolution \citep{p37, p38, p39, Yoon2006}. This is discussed in more detail in Section \ref{subsec:MOBresults}. 

In this paper, we use the results from detailed simulations of the CHE scenario as developed by \cite{p36} and combine these with the cosmological simulations of the chemical and star-formation history of the Universe by \citet{2015MNRAS.448.1835T} to conduct a binary population synthesis study and to investigate the population properties, cosmological rates and aLIGO detection rates of BHBHs. We also determine the cosmological pair-instability supernova (PISN) rate in this scenario. PISNe are rare events that occur when pair production in the core of a very massive star softens the equation of state causing it to become unstable before oxygen is ignited in its core. This leads to a partial collapse followed by a runaway thermonuclear explosion that blows the star apart, leaving no remnant \citep{FowlerHoyle1964,RakaviShaviv1967}. This is expected to happen to low-metallicity stars with a final helium core mass $M_\textrm{He,f}$ in the range of \mbox{$60 \lesssim  M_\textrm{He,f}/M_\odot \lesssim 125$} (\citealt{p4,weasle2017,marchant2018}). As there are no compact remnants in this range predicted by binary evolution models, and hence no potential BHBH mergers from that pathway either, we refer to this as the PISN gap\footnote{A BHBH merger with one or both components in the PISN mass gap, GW190521 \citep{GW190521}, was observed while this paper was under review; as discussed by \citet{GW190521:astro}, it may have originated from other evolutionary channels.}. These events are also of interest as they can be used to probe stellar evolution in low-metallicity regimes.

We find cosmological BHBH merger rates through the CHE channel that may represent a non-negligible fraction of the total merger rate \citep{review2019} and show that we might soon be detecting massive mergers from above the PISN gap. In addition, the presence of systems with very small delay times may make it possible to use future detections of high-redshift mergers to provide insight to the evolution of massive stars in the early Universe.

The paper is structured as follows: Section \ref{subsec:SFR} gives an overview of the hydrodynamical simulations and resulting star-formation history of \citet{2015MNRAS.448.1835T} that is used in our simulations, Section \ref{subsec:MOBresults} describes the work of \cite{p36} and the binary calculations used in this study, and Section \ref{subsec:mc} summarizes our simulations, the overall setup and the assumptions made. Section \ref{sec:results} presents our main results: final population properties and rates, the effects of possible momentum kicks during black-hole formation, the outcomes of deviations in the high-redshift SFR and the calculation of PISN rates and their detectability. Lastly, Section \ref{sec:conclusion} gives concluding remarks and discusses future work.


\section{Methods}
\label{sec:methods}

\subsection{Star-formation history and metallicity distribution}
\label{subsec:SFR}
As was shown by \cite{p36}, the properties of the BHBH mergers formed through the CHE channel, e.g. the final black-hole masses and the delay times, are a strong function of the metallicity\footnote{Metallicity is defined as the mass fraction of elements heavier than helium.} of the progenitors. Therefore, to estimate the merger rates at the current epoch (or any other epoch), one needs not only the star-formation history of the Universe as a function of redshift but also its dependence on metallicity (see also \citealt{Chruslinska2019a}, \citealt{Chruslinska2019b}, \citealt{Belczynski2017} and \citealt{Neijssel2019} on the uncertainties in the metallicity-specific star-formation rate). We obtain these distributions from the cosmological simulation by \citet{2015MNRAS.448.1835T} -- a self-consistent, hydrodynamical simulation in a $(25 h^{-1} \textrm{Mpc})^3$ volume that includes star formation, feedback from supernovae, active galactic nuclei \citep{2014MNRAS.442.2751T}, and the effects of chemical enrichment. Here $h$ is the dimensionless Hubble constant, taken to be $0.7$. The resolution of simulated galaxies is $1.4 \times 10^7 h^{-1} \textrm{M}_{\odot}$, meaning that the smallest reliable galaxies would lie around $\sim 10^9 \textrm{M}_{\odot}$ and that galaxies below this resolution are not constrained from observations. However, the contributions of such low-mass galaxies to stellar mass and metallicities are small, and thus we include all particles of the simulation volume in the analysis of this paper. Note that, however, given the extremely low metallicity of dwarf galaxies, they can be important environments for the formation of massive black hole progenitors (see e.g. \citealt{toffano2019}).  If the metallicities of such dwarf galaxies in the simulations are overestimated by $>> 0.5$ dex (due to non-universal nucleosynthesis yields or the initial mass function), then it could affect the rates determined in this paper.

This simulation uses the most recent chemical enrichment sources: the nucleosynthesis yields are taken from \citet{2011MNRAS.414.3231K} (see also \citealt{Nomoto2013}), which are consistent with the elemental abundances observed in the Milky Way\footnote{While this paper was in review, new nucleosynthesis yields from \cite{2020chiaki} became available - we estimate that this would lead to a roughly $20\%$ increase in the BHBH rates reported in this paper.}. The modeling of Type Ia supernova progenitors here is based on the single degenerate scenario and includes metallicity effects \citep{2009ApJ...707.1466K}, which affect the timescale for iron enrichment. The simulation assumes a Kroupa initial mass function (IMF) for all galaxies \citep{kroupa2001}.
This simulation can also reproduce various observations of galaxies: the so-called $M$--$\sigma$ relation\footnote{The correlation between the stellar velocity dispersion $\sigma$ in the bulge of a galaxy and the mass $M$ of the supermassive BH at its centre.}, the black hole mass to bulge mass relation, the galaxy size to mass relation and, most importantly for our study, the stellar and gas-phase mass-metallicity relations \citep{2015MNRAS.448.1835T} as the black-hole merger production will depend crucially on these relations. In the simulation, AGN cause large-scale metal-enriched outflows \citep{2015MNRAS.452L..59T}, the mass-metallicity relation evolves as a function of time \citep{taylor2016}, and the metallicity radial gradients loosely correlate with the galaxy mass \citep{2017MNRAS.471.3856T}.

\begin{figure}
\centering
\includegraphics[width=0.45\textwidth]{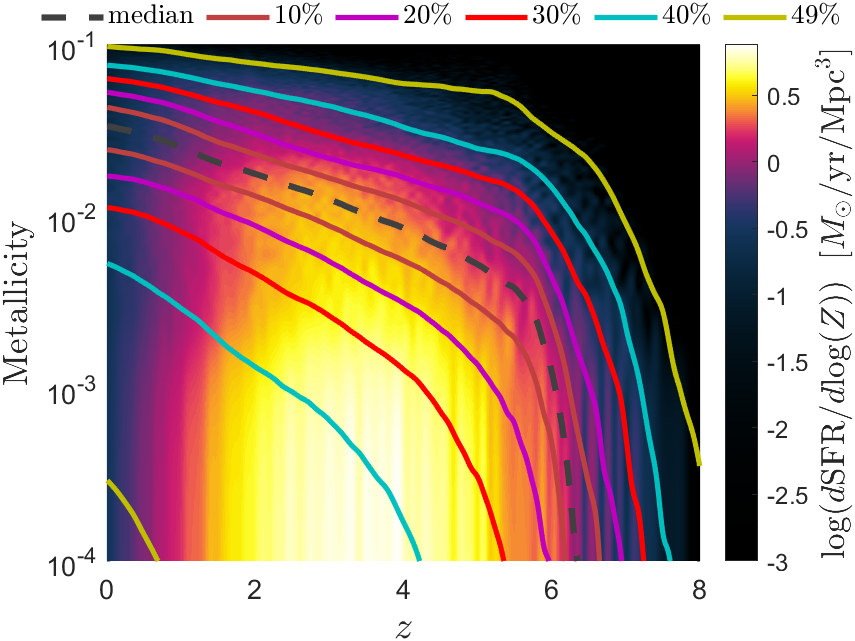}
\caption{The density distribution of the star-formation rate (SFR) as a function of metallicity ($Z$) and redshift ($z$), obtained from the hydrodynamical simulations of \citet{2015MNRAS.448.1835T}, described in Section \ref{subsec:SFR}. The curves show quantiles of the cumulative metallicity distribution of star formation at a given redshift. One can clearly see that the median metallicity increases towards low redshifts and that the SFR becomes insignificant at high redshifts of the order of  $z \sim 8$. This distribution provides the statistical formation redshift and metallicity input to our CHE modelling.}
\label{fig:chiaki3}
\end{figure}

\begin{figure}
\centering
\includegraphics[width=0.45\textwidth]{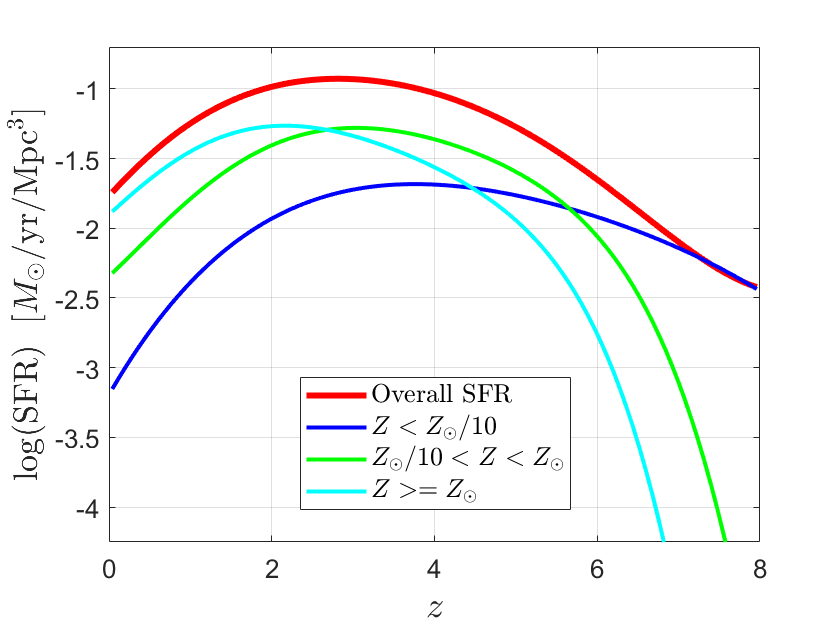}
\caption{The Madau plot shown for different ranges of metallicity for the hydrodynamical simulations of \citet{2015MNRAS.448.1835T}, discussed in Section \ref{subsec:SFR}. \textbf{Red:} The Madau plot for all ranges of metallicity; \mbox{\textbf{Dark blue:}} For metallicities $Z < Z_{\odot}/10$; \textbf{Green:} $Z_{\odot}/10 < Z < Z_{\odot}$; \textbf{Light blue:} $Z > Z_{\odot}$. For this work, solar metallicity $Z_{\odot}$ is taken to be 0.017 as in \citet{in2}. For the metallicity range considered in our Monte-Carlo simulation ($-5.0 < \log(Z) < -2.375$, as discussed in Section \ref{subsec:mc}), the dark blue and green curves are the only relevant curves.}
\label{fig:chiaki2}
\end{figure}

The simulation tracks the formation redshift and metallicity at the formation time for all stars that have formed up to redshift zero, with cosmological parameters $\Omega_M = 0.28$ and $\Omega_{\Lambda} = 0.72$ \citep{hinshaw2013}. We can therefore track the SFR as a function of both redshift and metallicity. In order to obtain a figure of merit that will allow us to draw from the SFR distribution as a joint function of redshift and metallicity, we produce a two-dimensional matrix displayed in Figure \ref{fig:chiaki3}, showing the differential star-formation rate per unit metallicity. Also shown in the figure are the percentiles for the SFR, given a certain redshift and metallicity. By integrating this joint function over all metallicities the usual Madau plot is obtained (\citealt{madau2014}): Figure \ref{fig:chiaki2} shows the Madau plot for different metallicity ranges, as well as the overall SFR in red, with the Solar metallicity taken as $Z_{\odot}=0.017$ \citep{in2}. The expectation that the metallicites are modelled correctly and self consistently allow us to use the resulting SFR distribution to model low-metallicity binaries at all redshifts --- these are the systems acting as possible BHBH progenitors in our Monte Carlo simulations. By integrating the SFR over the observable Universe, we can calculate the total mass converted into stars as $2.4 \times 10^{20} M_\odot$.


\subsection{Results from detailed CHE evolution models}
\label{subsec:MOBresults}

The CHE scenario requires binary systems that are in near contact at the onset of hydrogen burning. The tidal interactions between the components of one such binary system can cause rapid stellar rotation leading to the deformation and deviation from spherical symmetry, which for massive stars (where the effects of radiation pressure are important) have been shown to keep the stars chemically mixed and homogeneous throughout core hydrogen burning \citep{eddington1924, p37, p38, p39}. For single stars, low metallicity is required to ensure that mass loss from stellar winds is low to avoid the associated spin-down, so that the stars can essentially remain chemically homogeneous up to core hydrogen exhaustion \citep{p40, p41, p43, p42, p44}. Due to their lack of composition gradients, these stars do not maintain a hydrogen-rich envelope and as such avoid the normal dramatic expansion during the post-main sequence phase. This motivated \cite{p34} to propose that close binary systems consisting of rapidly rotating, massive stars at low metallicities undergoing chemically homogeneous evolution could avoid merging during the early evolution and evolve without ever undergoing a contact or mass-transfer phase (see also \citealt{p35, p1, p45}).

In contrast, through detailed binary evolution models where the efficiency of rotational mixing processes had been calibrated to match the range of nitrogen abundances observed in early type stars \citep{brott2011}, \cite{p36} found that contact-free evolution in these close binaries is very rare, and that many systems undergo contact (referred to as 'overcontact') but still avoid merging during the core hydrogen burning phase. During overcontact, both stellar components overfill their Roche volumes. The evolution during the overcontact phase differs from the standard evolution during a more classical CE phase as co-rotation is maintained as long as material stays inside the L2 point. The binary system can therefore avoid a spiral-in phase due to viscous drag, giving rise to a stable system that evolves on a nuclear timescale. If both stars overflow the L2 point, however, the system is expected to rapidly merge either due to the resulting angular-momentum losses or due to an inspiral caused by a loss of co-rotation.


A system will typically enter an overcontact phase at the early stages of core hydrogen burning. Mass will be transferred back and forth between the binary components until their masses are roughly equal. Mass loss from stellar winds is expected to result in a widening of the orbit, driving the system out of contact. As these winds are metallicity dependent \citep{Vink2001, p47, p46}, more metal-rich systems evolve to longer orbital periods, and as the system remains tidally locked its rotational velocity can be lowered to the point where chemically homogeneous evolution is no longer effective. Even if the system remains chemically homogeneous, the orbit may be too wide for the resulting BHBH  system to merge in a Hubble time. This is the reason why the CHE scenario, at least in its simplest version, requires low metallicity.

The binary calculations used in this study are based on the CHE scenario as described by \cite{p36}, but extended until core carbon depletion and covering a much larger range of initial conditions. Using the MESA v11701 stellar evolution code \citep{paxton2011, paxton2013, paxton2015, paxton2018, paxton2019}, we extended these calculations to obtain a fine grid of models as a function of metallicity, initial primary mass and initial orbital period for a fixed initial mass ratio \mbox{$q_\textrm{i} = M_{2,\textrm{i}}/M_{1,\textrm{i}} = 1$}. The metallicity range was taken as $-5.0 \leq \log(Z) \leq -2.375$, where \mbox{$-2.375 \hspace{1mm} \approx \log(Z_{\odot}/3)$)}. The initial primary mass range was taken as \mbox{$25 \leq M_{1,\textrm{i}}/M_\odot \leq 500$}, and the initial orbital period range was taken as $0.4 \leq P_\textrm{i}/{\rm d} \leq 4.0$ (see Appendix \ref{AppA} for details). In total, we calculated roughly 150,000 new binary evolution sequences (as compared to about 2,000 used in \citealt{p36})\footnote{The results of our MESA simulations, including tables summarizing the outcomes and the necessary input files to reproduce them, are available at \url{https://doi.org/10.5281/zenodo.3348337}. The source code used for our Monte Carlo calculations, as well as the results of our Monte Carlo simulations, are provided here as well.}.


To account for the occurrence of PISNe, we use the results from the hydrodynamical simulations of \cite{MarchantPPISN} who find that stars that deplete core carbon with helium core masses between $60.8 <M_{\textrm{He,f}}/M_\odot < 124$ are completely disrupted and leave no remnant. Stars just below the PISNe threshold are also expected to become unstable due to pair-creation in their cores, but the resulting thermonuclear explosion only removes a fraction of the total mass. The star can then undergo multiple pulsations and associated mass loss events before finally collapsing to a BH, a process which is called a pulsational-PISN (PPISN, \citealt{Fraley1968,weasle2017}). To account for mass loss before BH formation from PPISNe we also consider the results of \cite{MarchantPPISN} who find this outcome to happen for stars with helium core masses between $35.1 <M_{\textrm{He,f}}/M_\odot < 60.8$ at core carbon depletion (see also \citealt{Farmer2019}). For stars falling in this mass range we interpolate the final BH masses predicted by \cite{MarchantPPISN}. The impact on the orbit of mass ejected through a PPISN is modeled as a single Blaauw kick \citep{Blaauw1961}, which modifies the semi-major axis of the binary and its eccentricity. Afterwards, we assume that the system circularizes while conserving orbital angular momentum, resulting in an orbital separation $a=a_0(1-e_0^2)$ where $a_0$ and $e_0$ are the semi-major axis and eccentricity resulting from the Blaauw kick. This is motivated by the results of \cite{MarchantPPISN}, who find that PPISNe occurring in binary systems compact enough to produce a merging BHBH would lead to Roche-lobe overflow before BH formation. Stars that deplete carbon with $M_{\textrm{He,f}} <35.1 M_\odot$ or $M_{\textrm{He,f}} >124 M_\odot$ are assumed to directly collapse into a BH with no mass loss.

\begin{figure}
\centering
\includegraphics[width=0.47\textwidth]{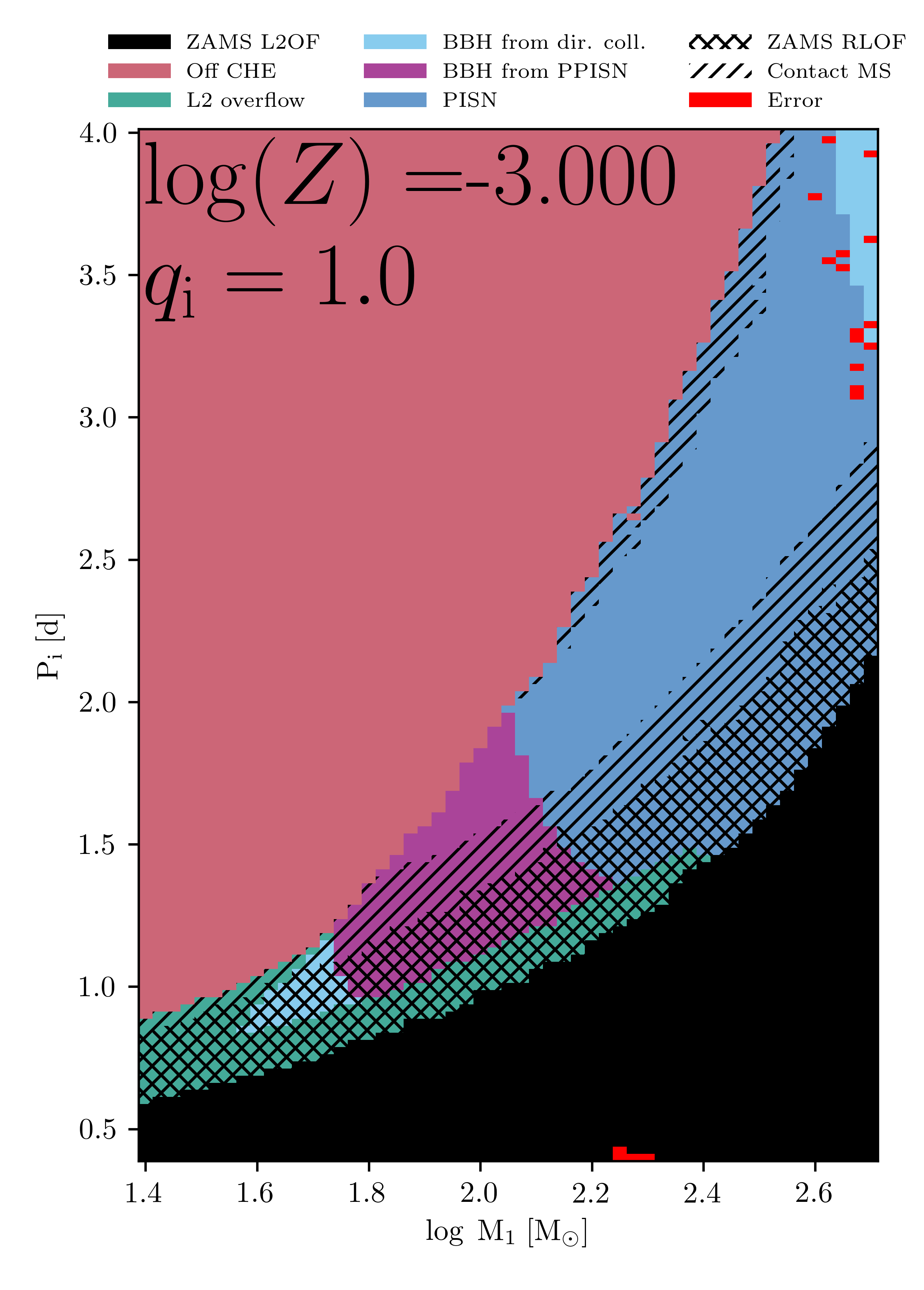}
\caption{An example of a grid of binary systems showing their initial period $P_\textrm{i}$ (in days) and initial primary mass $M_{1,\textrm{i}}$ for $\log(Z)  = -3$ and $q_\textrm{i} = 1$, with their final outcomes colour-coded according to the legend at the top. Black regions have initial periods small enough to have L2 overflow at the ZAMS, while green areas are systems reaching L2 overflow during the main sequence -- for both of these scenarios, systems rapidly merge. Pink areas depict systems where rotational mixing is not efficient enough and where at least one component no longer evolves chemically homogeneously; these systems are thus not considered progenitors of BHBH systems in this CHE study. Blue regions successfully form double He star binaries that will either collapse to form BHBH systems (light blue) or whose members will explode in pair-instability supernovae (dark blue) at a later stage. Purple denotes areas forming BHBH systems via the PPISN path. Single hatch-marks show systems experiencing contact during the main sequence, while double hatch-marks are models in contact at the ZAMS already. Red regions show models for which the simulations did not converge. For more information and examples of grids, see \mbox{Appendix \ref{AppA}.} (ZAMS: Zero Age Main Sequence).}
\label{fig:textGrid}
\end{figure}

An example of one of our model grids for $\log(Z) = -3$ is shown  in Fig. \ref{fig:textGrid}, with each rectangle corresponding to one detailed binary evolution model and showing the final fate of each system (the final fate being a function of the final calculated binary parameters). Possible fates include systems that fail to evolve chemically homogeneously, systems that have L2 overflow at the ZAMS or on the main sequence and therefore rapidly merge long before collapsing into black holes, and systems that successfully form double He stars (see Fig. \ref{fig:textGrid} for more information). We are interested in the latter group, as they will either collapse to form BHBHs or become pair-unstable. It can be seen that, for double massive He stars to be the final outcome, initial primary masses above $\sim 40 \hspace{1mm} \text{M}_{\odot}$ are required, and that the range of initial periods of progenitor systems for these binaries widens with an increase in primary mass. It is worth noting that, for most binaries below the PISN regime, overcontact systems dominate the progenitor space, and that contact-free evolution is only really found in systems having initial primary masses above $100 \hspace{1mm} \text{M}_{\odot}$. See Appendix \ref{AppA} for further examples of the grids used.



\subsection{Monte Carlo simulations}
\label{subsec:mc}
We systematically explore CHE evolution by carrying out Monte Carlo simulations that draw massive binary systems from a co-moving box of fixed co-moving volume\footnote{The co-moving volume is defined as the volume measure for which the number density of objects locked into the Hubble flow remains constant with redshift.} $V_{\textrm{MC}}$ stretching from the birth of the Universe to $z=0$. We then transform the rates calculated from this to rates observable from Earth by integrating over our light cone. This Section first describes the general outline of the simulation and the way in which binaries are sampled, whereafter a discussion of the rates and their normalisation follows.

\subsubsection{Sampling}
\label{subsubsec:mc_general}
The Monte Carlo method is used to draw the birth redshift ($z_0$), metallicity ($Z_0$), initial orbital period ($P_{\textrm{i}}$) and initial primary mass ($M_{1,\textrm{i}}$) of each binary system over parameter ranges corresponding to that of our binary evolution model grids (see Section \ref{subsec:MOBresults}).

In order to sample binaries from the required ranges of metallicity ($-5.0 \leq \log(Z) \leq -2.375$) and redshift ($0 < z < 8$), we produce a sampler that samples birth redshifts $z_0$ and metallicities $Z_0$ via rejection sampling, as described in Appendix \ref{AppB}. Initial primary masses were obtained by sampling a Salpeter initial mass function (IMF) over the range \mbox{$25 \leq M_{1,\textrm{i}}/M_\odot \leq 500$} \citep{salpeter1955}, and the mass ratio $q_\textrm{i} = M_{2,\textrm{i}}/M_{1,\textrm{i}}$ is assumed to be 1. The initial orbital period $P_\textrm{i}$ is drawn from a flat distribution in $\log(P_\textrm{i})$ over the range $0.4 \leq P_\textrm{i}/{\rm d} \leq 4.0$ (see also \citealt{p36}). We sample a billion binaries formed within these reduced parameter ranges in a co-moving volume\footnote{See Eq. \ref{eq:Vmc_last} and Section \ref{AppB_1} for the calculation of the co-moving volume $V_{\textrm{MC}}$.} of $V_{\textrm{MC}} = 1.90 \times 10^5 \hspace{1mm} \textrm{Mpc}^3$.

\subsubsection{Fate and final parameter values}
\label{subsubsec:fate}
Once the initial parameters for a given binary are determined, we perform a series of steps to predict first the fate and thereafter the final parameter values of the system. For this, we use our grid of detailed binary evolution models. This grid has three dimensions: one for metallicity, one for the initial primary mass and one for the initial orbital period for the binary, for the ranges listed in Section \ref{subsec:MOBresults} (and corresponding to the ranges over which our initial parameters are sampled - see Section \ref{subsubsec:mc_general}). Each point in the grid has a corresponding final fate assigned to it (as the evolution for each point was followed, where possible, up to core carbon depletion, whereafter the fate of the binary was taken to be determined - see Section \ref{subsec:MOBresults}).

Possible fates include systems that fail to evolve chemically homogeneously, systems that have L2 overflow at the ZAMS or on the main sequence and therefore rapidly merge long before collapsing into black holes, and lastly those systems that successfully form double He stars --- these include systems that collapse directly into BH binaries, systems that undergo pulsational mass loss before collapsing into BH binaries, and systems that end as PISNe (see \mbox{Figure \ref{fig:textGrid}} for a slice of the grid at $\log(Z) = -3.0$ and a description of the various model outcomes).

We use a weighted nearest-neighbour method to determine the fate (or `class') of each new Monte Carlo binary using our grid. For each of the 8 grid points surrounding a binary, its weight $w$ is calculated as $w = e^{-d^2}$, where $d$ is the 3-dimensional grid distance between the binary and the grid point under consideration (the 3 dimensions of the grid have units of $\log(\textrm{M}/M_{\odot})$, $\log(Z)$ and days for this calculation). The weights from grid points of the same class are added, and the binary is assigned the class corresponding to the greatest cumulative weight. We note here that, due to practical reasons, not all data points in the grids gave converged results, mainly because the highest metallicities and highest masses lead to simulation complexities resulting in unreasonably long run times (see the red areas in Figure \ref{fig:textGrid}). These points were treated as their own class by the nearest-neighbours method.

Once a binary system is classified as eventually either directly collapsing into a BH binary (`direct collapser'), undergoing pulsational mass loss before collapsing (`PPISN system') or resulting in a PISN (`PISN system'), our next step is to determine the final mass ($M_\textrm{He,f}$) of its components and the final orbital period ($P_\textrm{f}$) at core carbon depletion. Both the primary and secondary component in the binary are taken to have the same final mass $(M_{1,\textrm{He,f}} = M_{2,\textrm{He,f}} = M_{\textrm{He,f}})$. Our grid provides these values for each grid point corresponding to one of these three classes. Additionally, for grid points classified as PPISN systems, the grid supplies the final mass and period parameter values for both the case in which we ignore PPISN mass loss, as well as the case where we include it --- this gives us the option of either including or excluding the effects of pulsational mass loss in our binary population synthesis. Final mass and period values for our binaries are obtained by applying a B-spline kernel\footnote{The B-spline kernel is of 3\textsuperscript{rd} order in $M_{1,\textrm{i}}$ and $P_\textrm{i}$ and 2\textsuperscript{nd} order in metallicity.} to the surrounding grid points of similar classes. As we assume that systems essentially remain the same after core carbon depletion, these are the masses and periods direct collapsers and PPISN systems will have after collapsing to form BHBH binaries, and it is also the masses and periods of PISN systems at the point where they become pair-unstable and explode.
ii

\subsubsection{BHBH and PISN calculations}
\label{subsubsec:calcs}
For direct collapsers and PPISN systems, we next calculate the delay time $t_\textrm{d}$ of the system in order to determine whether it will merge at the present epoch and as such contribute to the present merger rates. Delay time is here defined as the time passing between the system's collapse into a BHBH system and its eventual merger. For our standard model, we assume that regular core-collapse supernovae allow He stars to collapse into black holes without any mass loss or kicks \citep{Fryer1999}, but incorporate mass loss (though not kicks) for PPISNe, as described below. The delay time is then calculated using \cite{Peters1964}:

\begin{equation}
t_\textrm{d} = \frac{5 a^4 c^5}{256 \mu M_\textrm{b}^2 G^3}.
\label{eq:time_delay}
\end{equation}

\noindent Here $M_\textrm{b}$ is the sum of the two black hole masses, given by \mbox{$M_\textrm{b} = 2 M_\textrm{He,f}$}, $\mu$ is the reduced mass of the system, which here is just $\mu = M_\textrm{He,f}/2$, $G$ is the gravitational constant, $a$ is the orbital separation and $c$ is the speed of light. Once the delay time is determined, the merging redshift $z_m$ of the binary can be calculated.

If, on the other hand, a binary is classified as a PISN system, we estimate each component's Ni mass $M_\textrm{Ni}$ by making use of a power law fit in log to the results of \cite{p4}, whose He mass range spans $65-130 \hspace{0.5mm} M_{\odot}$:

\begin{equation}
\log(M_\textrm{Ni}/M_{\odot}) = r (M_\textrm{He,f}/M_{\odot})^s + t.
\label{eq:heger_fit}
\end{equation}

\noindent Here $r = -5.02 \times 10^4$, $s = -2.179$ and $t = 2.887$. From this, we next calculate the peak bolometric magnitude due to Ni decay $M_{\textrm{bol,Ni}}^{\textrm{peak}}$ with the use of an Arnett-like relation \citep{Arnett1982}

\begin{equation}
M_{\textrm{bol,Ni}}^{\textrm{peak}} = -19.2 - 2.5 \log \left(\frac{M_\textrm{Ni}}{0.6 M_\odot}\right),
\label{eq:arnett}
\end{equation}

\noindent from which the bolometric peak luminosity $L_{\textrm{bol,Ni}}^{\textrm{peak}}$ can be obtained from

\begin{equation}
\log\bigg(\frac{L_{\textrm{bol,Ni}}^{\textrm{peak}}}{L_\odot}\bigg) = \frac{M_{\textrm{bol},\odot} - M_{\textrm{bol,Ni}}^{\textrm{peak}}}{2.5},
\label{eq:lum_Ni}
\end{equation}

\noindent where $M_{\textrm{bol},\odot}$ is the solar bolometric magnitude (taken as 4.75 - see \citealt{Casagrande2006}) and $L_\odot$ the solar luminosity. Low-mass PISNe produce very little Ni \mbox{($M_{\textrm{Ni}} \sim 2 \times 10^{-4} M_{\odot}$ for the lowest-mass PISNe)}, such that the luminosity of the shock heating dominates; the peak luminosity of the PISN is then determined by the explosion energy of the PISN instead of its Ni mass. From \cite{Herzig1990} it can be seen that the peak luminosity of a $61 \text{M}_\odot$ He core SN is dominated by the explosion energy, and from \cite{Kasen2011} it is evident that for a $70 \text{M}_\odot$ SN, the peak luminosity is still only barely powered by Ni. For our calculations we make the assumption that the turning point occurs at roughly $70.5 \text{M}_\odot$ (where the explosion magnitude and the magnitude resulting from Ni decay are roughly equivalent), and with the help of Equations \ref{eq:heger_fit} and \ref{eq:arnett} we then determine the explosion peak bolometric magnitude $M_{\textrm{bol,exp}}^{\textrm{peak}}$ to be $-15.2$ and convert that to the equivalent luminosity $L_{\textrm{bol,exp}}^{\textrm{peak}}$ using the form of Equation \ref{eq:lum_Ni}. We further assume that this explosion luminosity is constant for all PISNe, regardless of He mass. As this explosion magnitude is faint and only dominates the peak luminosity for the faintest PISNe, the effect of this assumption will be negligible for brighter PISNe and any magnitude limited results should be relatively insensitive to this. The total peak luminosity $L_{\textrm{bol,tot}}^{\textrm{peak}}$ of a PISN system is then just $L_{\textrm{bol,Ni}}^{\textrm{peak}} + L_{\textrm{bol,exp}}^{\textrm{peak}}$.



Obtaining the maximum absolute bolometric magnitude $M_{\textrm{bol}}^{\textrm{peak}}$ from $L_{\textrm{bol,tot}}^{\textrm{peak}}$ from Equation \ref{eq:lum_Ni}, using the luminosity distance $d_\textrm{L}$ of the PISN system and ignoring any obscuration effects, it is then possible to calculate its maximum apparent bolometric magnitude $m_{\textrm{bol}}^{\textrm{peak}}$ as

\begin{equation}
m_{\textrm{bol}}^{\textrm{peak}} = M_{\textrm{bol}}^{\textrm{peak}} + 5 \log \left( \frac{d_\textrm{L}}{10\text{pc}} \right) ,
\label{eq:apparent}
\end{equation}

\noindent which can then be used to determine whether a specific system will be detectable with a magnitude limited survey or not. These statistics can then be obtained to estimate the observable PISN rate.

\begin{figure}
\centering
\includegraphics[width=0.45\textwidth]{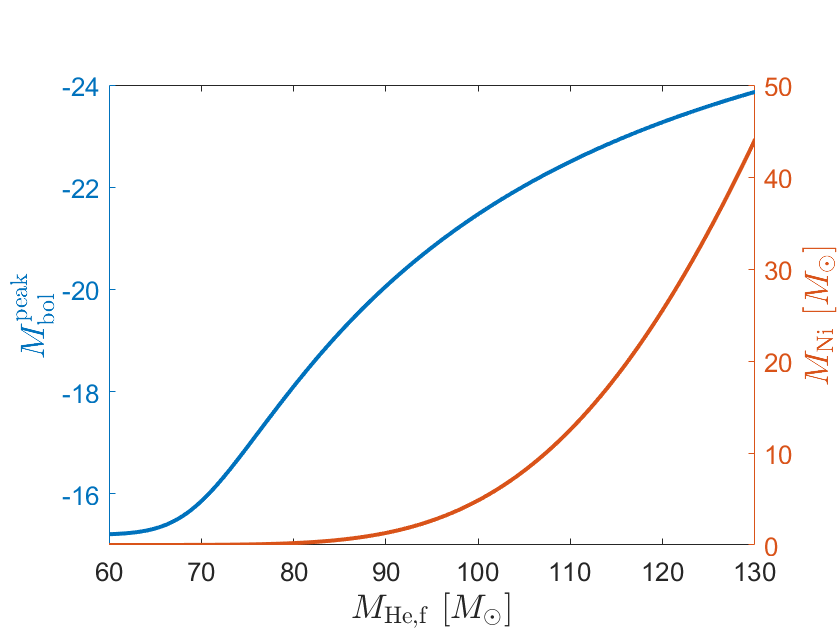}
\caption{The relation between the final He mass $M_{\textrm{He,f}}$ of a star and the absolute peak bolometric magnitude $M_{\textrm{bol}}^{\textrm{peak}}$ associated with its PISN, as determined by Equations \ref{eq:heger_fit} and \ref{eq:arnett}, and corrected as described below \mbox{Equation \ref{eq:lum_Ni}}. The relation between $M_{\textrm{He,f}}$ and the Ni mass $M_{\textrm{Ni}}$ of a star is also shown. We further note that the calculations done for this figure assume no extinction.}
\label{fig:PISN_absVShe}
\end{figure}

Figure \ref{fig:PISN_absVShe} shows the relation between the final He mass $M_{\textrm{He,f}}$ of a star and the absolute peak bolometric magnitude $M_{\textrm{bol}}^{\textrm{peak}}$ associated with its PISN, as determined  by Equations \ref{eq:heger_fit} and \ref{eq:arnett}. We note that we do not apply K corrections that would tend to reduce the apparent magnitude at large redshifts. This therefore leads to an overestimate of the observable PISN rate at large redshifts.

\subsubsection{Rates and Monte Carlo normalisation}
\label{subsubsec:rates_norm}
We calculate BHBH merger rates as detected by the O1, O3 and full design aLIGO sensitivities as well as by the Einstein Telescope's (ET) predicted sensitivity, and also determine the cosmological BHBH merger rate. We further determine the intrinsic PISN rate along with the PISN detection rates for various magnitude limited surveys. The mathematical background for our rate calculations and for the normalisation of our set of Monte Carlo draws can be found in Appendix \ref{AppB}. \mbox{Figure \ref{fig:aLIGOcurves}} shows the various strain sensitivities of aLIGO and ET.


\section{Results}
\label{sec:results}
The results of our Monte Carlo simulations are detailed in this section. More precisely, Section \ref{subsec:finalproperties} discusses the final BHBH population properties while Section \ref{subsec:mergerrates} presents the overall BHBH merger rates and aLIGO detectability. Section \ref{subsec:kicks} explores the effects of momentum kicks on the BHBH population, Section \ref{subsec:PISNe} discusses derived PISN population properties and Section \ref{subsec:high_redshift} provides scenarios in which the high-redshift star-formation rate deviates from our standard cosmological simulations and discusses the effects of that on the previously derived BHBH population properties. It should be noted that, although we have a default scenario, we altogether investigate four different configurations. The default case includes pulsational mass loss effects, and is denoted by \textit{PPISN}. The three other cases are as follows: that of including both pulsational mass loss and momentum kicks (\textit{PPISN + Kicks}), that of excluding pulsational mass loss (\textit{NON-PPISN}) and that of excluding pulsational mass loss while including momentum kicks (\textit{NON-PPISN + Kicks}). Whenever not specified, the default configuration is assumed. In this section, we consider both co-moving volumetric merger rates (denoted by $d^{2}N/dV_c/dt$) as a function of merger redshift (see Eq. \ref{eq:com_rate} and \ref{eq:aligo_com_rate}), as well as these volumetric rates integrated over our past light cone to obtain rates potentially observable from Earth, denoted by $R$ (see Eq. \ref{eq:rm_main}/\ref{eq:rm2_main}). Where the need arises to specifically distinguish between detectable and cosmological rates, different subscripts will be used. As examples, $d^{2}N_{m}/dV_c/dt$ would be the cosmological volumetric merger rate, and $R_{det}$ would be the detectable merger rate. We only consider BHBH mergers and PISNe that evolved through the CHE channel; this should be kept in mind for the BHBH merger and PISN rates presented here.

\subsection{Final population properties}
\label{subsec:finalproperties}
\begin{figure}
\centering
\includegraphics[width=0.45\textwidth]{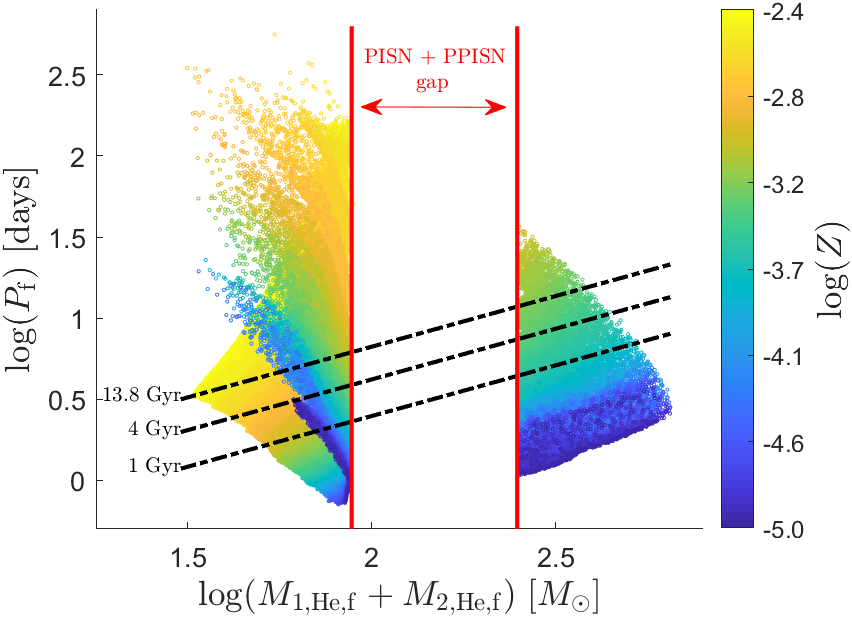}
\caption{The distribution of final periods and final combined masses of systems that reached core carbon depletion for the default (\textit{PPISN}) case. Note that the final He-core masses correspond to the final BH masses. The three black curves represent the delay time of mergers, with systems below the bottom curve merging in less than 1 Gyr, systems below the middle curve merging in less than 4 Gyr and systems below the topmost curve merging in less than 13.8 Gyr. The colouring of different points shows the metallicity (in $\log(Z) $) of the systems. Systems between the vertical red lines experience either PISNe or PPISN mass loss, and this area is referred to as the "PISN + PPISN gap", where no BHBHs form. Note that there is a significant number of very high-mass systems with very low metallicity merging in a Hubble time.}
\label{fig:heDepletion}
\end{figure}

Figure \ref{fig:heDepletion} shows the distribution of the total final binary mass and final orbital period $P_\textrm{f}$ of all the systems in our simulations that resulted in close He star pairs at the point of helium exhaustion. The range of final periods for systems at high metallicity is larger than that for low-metallicity systems due to stellar wind mass loss affecting systems of different masses differently at different metallicities. For systems at high metallicities with high initial masses, stellar winds will result in more mass loss and cause a more dramatic widening of the system, producing systems with longer orbital periods and lower final masses, while systems with lower initial masses at the same metallicity will be less affected. This explains the occurrence of systems with lower final masses and a larger final period range for high metallicities. On the other hand, for low metallicities, systems of high initial mass are much less affected by stellar wind mass loss (and lower masses are of course even less affected); systems can therefore retain more mass and avoid a significant widening of their orbits, which explains the larger range of final masses and smaller range of final periods for these metallicities. The upwards trend of final parameter values moving from higher final masses and lower final periods to lower final masses with longer final periods is caused by the inclusion of pulsational mass loss, whereby the mass loss causes a widening of the orbital period.

Figure \ref{fig:heDepletion} also indicates the merger times of the systems, assuming that all systems will collapse into black holes without any kicks or associated mass loss. All binaries falling above the 13.8 Gyr line will not be able to merge in a Hubble time.  At \mbox{$\log(Z)=-2.4$} there is a small parameter range where systems will merge in a Hubble time, while for most other systems at $Z > Z_\odot/10$, all but the lowest masses will not merge. When moving to lower metallicities, an increasing number of systems are able to produce short-period systems, and below $\log(Z) \sim -3.5$ most systems are able to merge in a Hubble time. This shows that the population of BHBH mergers should be dominated by low-metallicity populations. Note that the empty area between the two vertical red lines are due to systems resulting in PISNe \mbox{($60.8 \lesssim  M_\textrm{He,f}/M_\odot \lesssim 124$)} and systems that undergo PPISN mass loss, resulting in an effective ``PISN + PPISN gap" of \mbox{$43.9 \lesssim  M_\textrm{He,f}/M_\odot \lesssim 124$.}


\begin{figure}
\centering
\includegraphics[width=0.45\textwidth]{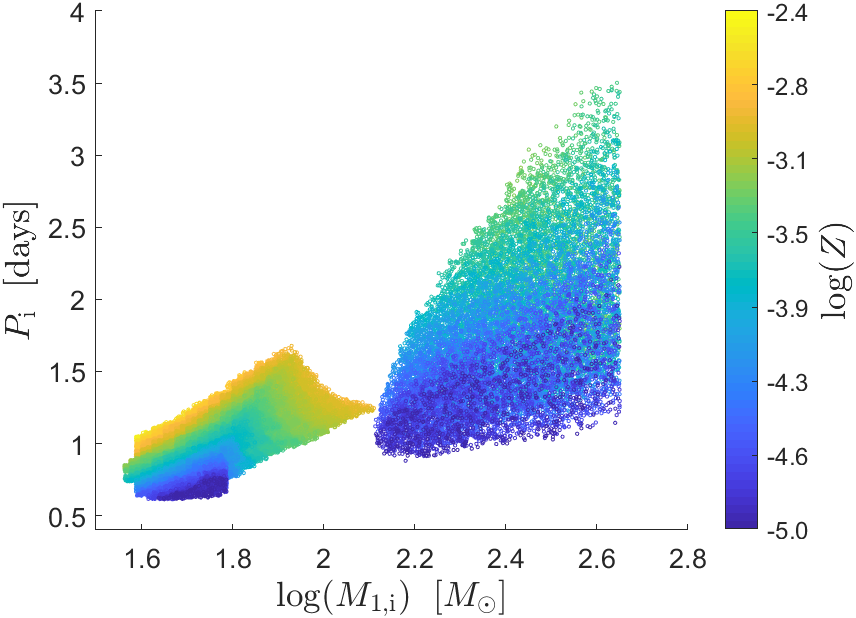}
\caption{The distribution of initial masses and initial periods for systems that will produce BHBH systems merging within the lifetime of the Universe, where the colour indicates the metallicity. The trend towards longer periods at higher masses can also be seen from the grids in Appendix \ref{AppA}.}
\label{fig:initial_P_vs_M}
\end{figure}

The initial periods and primary masses of systems eventually forming BHBH systems merging within the lifetime of the Universe are shown in Figure \ref{fig:initial_P_vs_M}, coloured according to their metallicity. It is clear that there are two progenitor populations -- one on either side of a gap caused by systems that eventually result in PISNe or undergo PPISN mass loss. For systems with high metallicities, the initial periods and primary masses fall in a narrow range, whereas lower metallicities give rise to a wider range of possible parameter values. This is due to the effects of stellar wind mass loss, where the effects at high metallicities are more severe and orbits can be widened more significantly, giving rise to only a small range of initial periods leading to BHBH systems with periods small enough for timely mergers. As the effects of winds increase with stellar mass, the successful higher-metallicity systems are constrained to smaller masses, in contrast to the systems at lower metallicities.

The cosmological and aLIGO full design sensitivity chirp mass\footnote{The chirp mass is defined as $M_{\textrm{chirp}} = (M_1 M_2)^{3/5} / (M_1 + M_2)^{1/5}$.} distributions of BHBH mergers merging in the lifetime of the Universe is shown in Figure \ref{fig:chirpmass} for both the \textit{PPISN} and \textit{NON-PPISN} case (see Appendix \ref{appB_detector} for the detector strain sensitivity curves and the calculation of the detection probability of mergers). The clear gap in the middle is due to PISNe for the \textit{NON-PPISN} case, and due to both PISNe and PPISN mass loss for the \textit{PPISN} case. For the \textit{PPISN} cases, the peaks in the distributions are due to PPISN pile-up, as discussed in \cite{Stevenson2019}. The mergers above the PISN gap are mostly comprised of very low-metallicity systems (as can be seen in \mbox{Figure \ref{fig:heDepletion})}, while those below the PISN gap also have contributions from higher metallicity systems. As the amplitude of gravitational waves resulting from the merger of two black holes is strongly dependent on the chirp mass, the smaller number of high chirp masses shown in the figure might still be a significant contributor to the BHBH detection rate, suggesting that very massive BHBH mergers may be detected from systems above the PISN gap. For our default \textit{PPISN} case, we find that eventual merger systems have BH masses in the range $17-43 \hspace{1mm} \textrm{M}_{\odot}$ below the PISN gap and $124-338 \hspace{1mm} \textrm{M}_{\odot}$ above the PISN gap (a range that includes intermediate-mass black holes). It should be noted that the upper mass of $338 \hspace{1mm} \textrm{M}_{\odot}$ is a function of the highest stellar masses (at $\sim 500 \hspace{1mm} \textrm{M}_{\odot}$) we consider.


\begin{figure}
\centering
\includegraphics[width=0.45\textwidth]{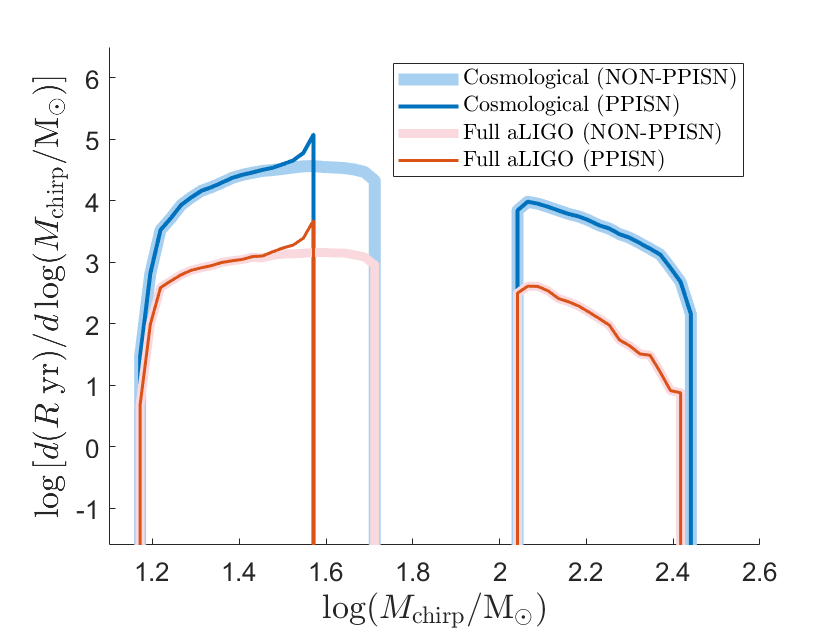}
\caption{The cosmological and aLIGO full design sensitivity chirp mass distribution of BHBH systems merging in the lifetime of the Universe for the \textit{PPISN} and \textit{NON-PPISN} cases.  $R$ denotes the BHBH merger rate observable from Earth (see Eq. \ref{eq:rm_main} and \ref{eq:rm2_main}). The total number of BHBH mergers in each of these four histograms correspond to their associated entries in Table \ref{tab:results}. The clear gap in the middle is caused by PISNe (for the \textit{NON-PPISN} case) and by PISNe and PPISN mass loss (for the \textit{PPISN} case).}
\label{fig:chirpmass}
\end{figure}

\begin{figure}
\centering
\includegraphics[width=0.45\textwidth]{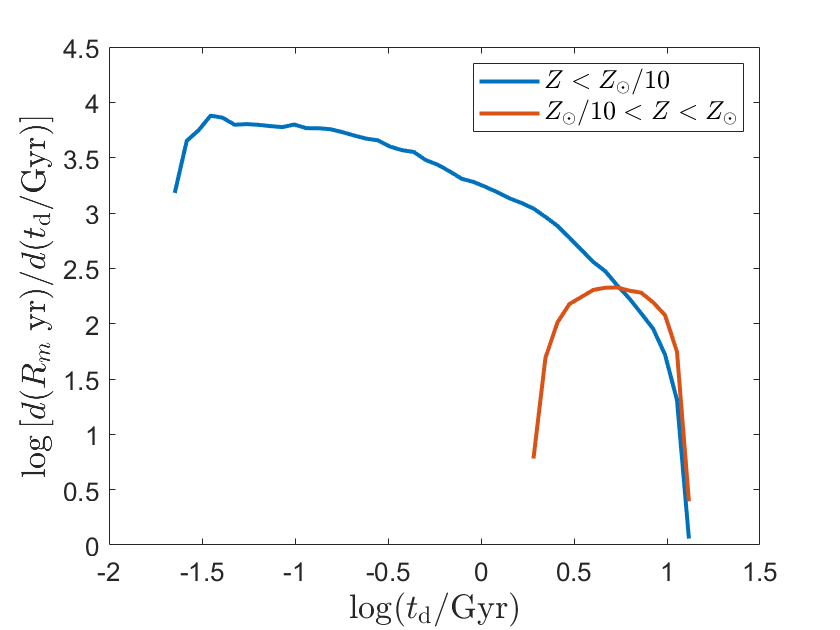}
\caption{The delay time ($t_{\textrm{d}}$) distribution of BHBH systems merging at the present epoch, with $R_m$ denoting the cosmological merger rate observable from Earth (see Eq. \ref{eq:rm_main} and \ref{eq:rm2_main}). Integrated over the delay time, the unit is therefore $\textrm{yr}^{-1}$. Mergers were divided into two metallicity ranges: \mbox{$Z < Z_{\odot}/10$} and $Z_{\odot}/10 < Z < Z_{\odot}$ (also see Figure \ref{fig:chiaki2}). These distributions have been normalised to the cosmological BHBH merger rate to give the total integrated BHBH merger rate shown in Table \ref{tab:results}.}
\label{fig:delay_times}
\end{figure}

Next, Figure \ref{fig:delay_times} shows the delay time distribution of BHBH systems merging in the lifetime of the Universe, showing a large range of delay times. Whereas \cite{p35} predict only delay times larger than 3.5 Gyr and \cite{p36} predict delay times no smaller than $\sim 0.4$ Gyr, our smallest delay time is $\sim 0.02$ Gyr. As Figure \ref{fig:heDepletion} shows, delay times systematically decrease with decreasing metallicity: at lower metallicities, the binary components are more compact and stellar winds affect them less, resulting in tighter BHBH systems with generally shorter delay times. \cite{p35} do not find this effect in their BHBH systems as they consider only one metallicity for all their binary models ($\log(Z) = -2.4$), near the threshold for chemically homogeneous evolution \citep{Yoon2006}, a metallicity that is close to the highest considered in this study ($\log(Z) = -2.375$). \cite{p36}, on the other hand, considered four different metallicities for their original detailed models, the smallest of which was $\log(Z) = -3.47$, and therefore found delay times much shorter than $3.5$ Gyr. As here we consider a continuous range of metallicities from $\log(Z) = -2.375$ all the way down to $\log(Z) = -5.0$, we are able to find even smaller delay times, pointing to the possibility of high-redshift merger detections with future detectors that are able to probe higher redshifts \citep{ET,CE}. The distribution of time delays will be highly dependent on the physical assumptions made. In the case of a flat initial $\log(P_{\textrm{i}})$ distribution and only gravitational radiation, we would expect a $1/t_{\rm d}$ distribution in time delay. However, here our time delay distribution depends both on gravitational radiation as well as the binary populations and their metallicity effects, the period distribution and stellar astrophysics.

\begin{figure}
\centering
\includegraphics[width=0.45\textwidth]{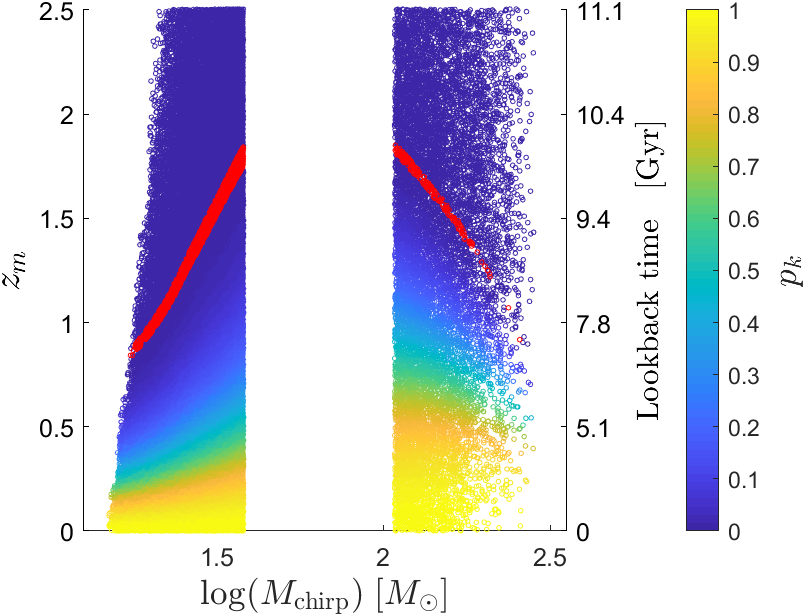}
\caption{The aLIGO detection probability $p_k$ at full design sensitivity for BHBH binaries as a function of the merger redshift $z_m$ and chirp mass $M_\textrm{chirp}$. The corresponding lookback time is given as a secondary y-axis. Note the rapid drop in detection probability as the redshift increases. Red points indicate those mergers for which $p_k = 0$; mergers above the red line therefore have no probability of detection.}
\label{fig:bhbhdetect}
\end{figure}

\begin{table*}
\centering
\caption{The cosmological BHBH merger rates (in units of $\textrm{yr}^{-1}$) and the BHBH merger detection rates for aLIGO at O1, O3 and full design sensitivity and for the planned Einstein telescope (ET) for both the \textit{PPISN} and \textit{NON-PPISN} cases, both with and without the inclusion of momentum kicks. Total merger rates are further broken down into mergers above and below the PISN gap.}
\label{tab:results}
    \begin{tabular}{@{}lcccccc}
    \hline
    ~        & ~              & Cosmological & aLIGO O1 & aLIGO O3 & Full aLIGO & ET \\ \hline
    ~        & Total          & 8510 & 16.8 & 80.6 & 310 & 7130 \\ 
    PPISN & Above PISN gap    & 1830 & 3.59 & 16.4 & 65.0 & 1220 \\
    ~        & Below PISN gap & 6680 & 13.2 & 64.2 & 245 & 5910 \\ \hline
    ~             & Total          & 1480 & 9.52 & 41.1 & 138 & 1340 \\
    PPISN + Kicks & Above PISN gap & 645 & 4.12 & 17.8 & 62.5 & 565 \\
    ~             & Below PISN gap & 839 & 5.40 & 23.3 & 75.0 & 770 \\ \hline
    ~            & Total          & 10100 & 20.0 & 98.3 & 382 & 8580 \\
    NON-PPISN    & Above PISN gap & 1830 & 3.59 & 16.4 & 65.0 & 1220 \\
    ~            & Below PISN gap & 8290 & 16.4 & 81.9 & 317 & 7360 \\ \hline
    ~        & Total              & 2070 & 12.8 & 57.0 & 193 & 1880 \\
    NON-PPISN + Kicks & Above PISN gap & 645 & 4.12 & 17.8 & 62.5 & 565 \\
    ~        & Below PISN gap      & 1420 & 8.68 & 39.2 & 131 & 1310 \\ \hline
     \end{tabular}
\end{table*}

\begin{figure*}
    \centering
    \begin{subfigure}[b]{0.35\textwidth}
        \includegraphics[width=\textwidth]{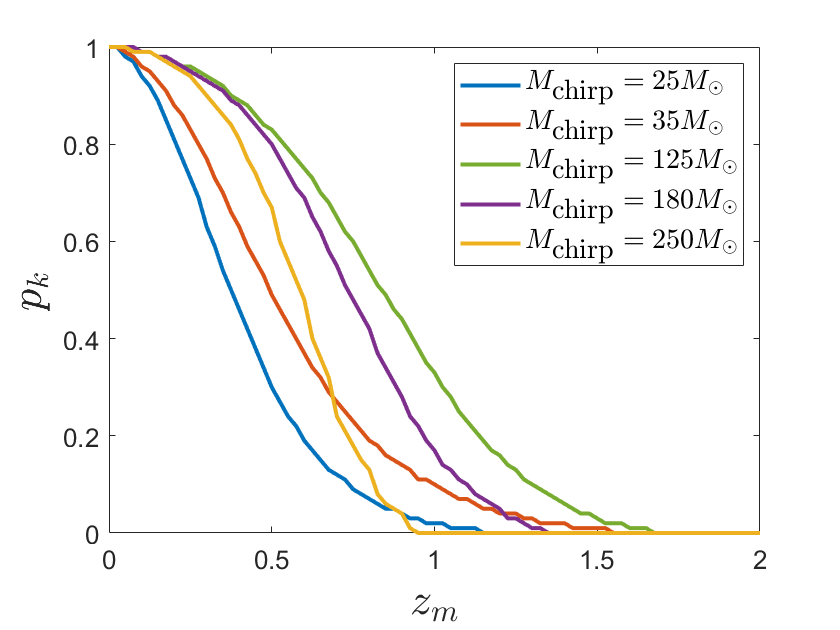}
        \caption{}
        \label{fig:full_sense_Mchirp}
    \end{subfigure}
    ~
    \hspace{-0.8cm}
    \begin{subfigure}[b]{0.35\textwidth}
        \includegraphics[width=\textwidth]{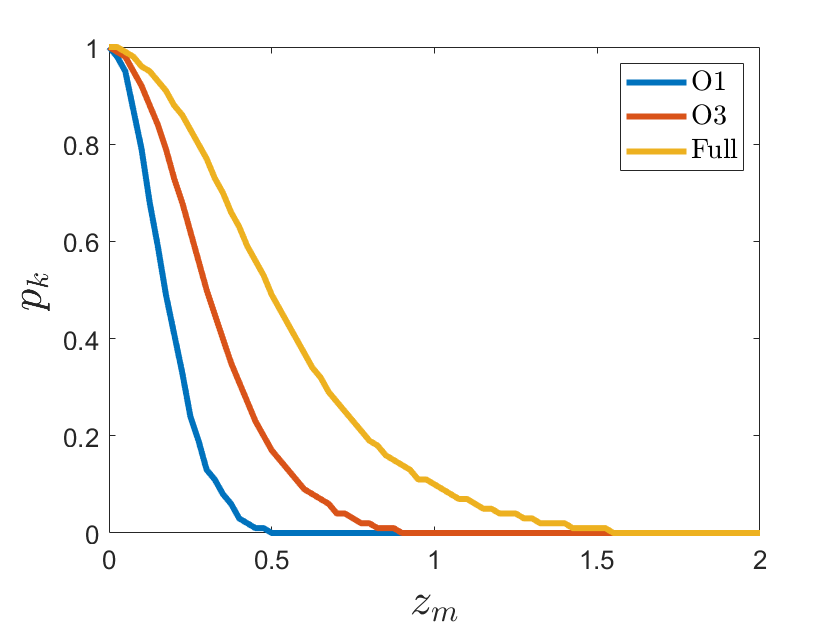}
        \caption{$M_\textrm{chirp} = 35 \hspace{1mm} \text{M}_{\odot}$}
        \label{fig:mchirp_35}
    \end{subfigure}
    ~
    \hspace{-0.8cm}
    \begin{subfigure}[b]{0.35\textwidth}
        \includegraphics[width=\textwidth]{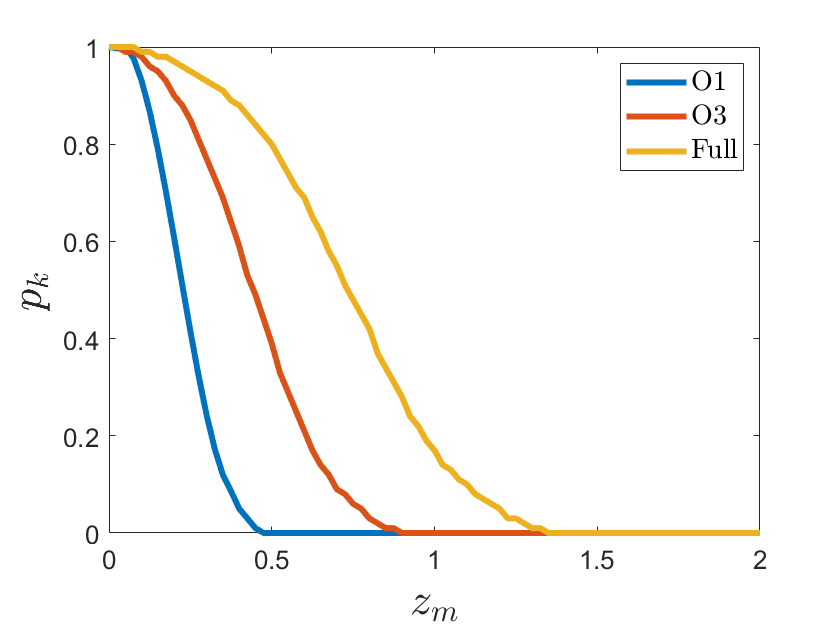}
        \caption{$M_\textrm{chirp} = 180 \hspace{1mm} \text{M}_{\odot}$}
        \label{fig:mchirp_180}
    \end{subfigure}
    \caption{\textbf{(a)} The aLIGO detection probability $p_k$ at full design sensitivity as a function of merger redshift $z_m$ for two mergers below the PISN gap ($M_\textrm{chirp} = 25 \hspace{1mm} \text{and} \ 35 \hspace{1mm} \text{M}_{\odot}$) and three systems above the PISN gap ($M_\textrm{chirp} = 125, 180 \hspace{1mm} \text{and} \hspace{1mm} 250 \hspace{1mm} \text{M}_{\odot}$). As Figure \ref{fig:bhbhdetect} also indicates, we expect no detections above a merger redshift of $z_m = 2$.  The most massive systems above the PISN gap, although easier to detect at lower merger redshifts, are less detectable (in comparison with systems below the PISN gap) at higher redshifts. This is due to the frequency sensitivity of aLIGO (see Figure \ref{fig:aLIGOcurves}). \textbf{(b)} The detection probability at full design, O1 and O3 aLIGO sensitivities as a function of merger redshift for a system below the PISN gap with chirp mass $M_\textrm{chirp} = 35 \hspace{1mm} \text{M}_{\odot}$. \textbf{(c)} Similar to \textbf{(b)}, but now for a system above the PISN gap with a chirp mass of $M_\textrm{chirp} = 180 \hspace{1mm} \text{M}_{\odot}$.}
    \label{fig:groupplot1}
\end{figure*}

\subsection{Black hole merger rates}
\label{subsec:mergerrates}
Using Appendices \ref{subsec:rates} and \ref{appB_detector}, we find the cosmological, O1, O3 and full design aLIGO sensitivity merger rates for all four of the configurations we investigate. These are summarised in Table \ref{tab:results}. As an interesting aside, in order to investigate potential future gravitational-wave detector capabilities, we decided to also inspect the predicted sensitivity of the Einstein Telescope\footnote{Einstein Telescope home: \url{http://www.et-gw.eu/}} (ET, \citealt{ET}), set to start operating in the next decade (see also the Cosmic Explorer\footnote{Cosmic Explorer home: \url{https://cosmicexplorer.org/}}, \citealt{CosmicExplorer}). Figure \ref{fig:aLIGOcurves} shows the most recent predicted sensitivity of the ET, along with those of aLIGO, and Appendix \ref{appB_detector} details the calculation of the detection probability of mergers. The ET is expected to find $83.8 \%$ of BHBH mergers whose signal reaches Earth at the present epoch, as is shown in Table \ref{tab:results}. At full aLIGO design sensitivity, \cite{p36} estimated the merger rate below the PISN gap to fall in the range $19 - 550\,\text{yr}^{-1}$, and that above the PISN gap to fall between $2.1 - 370\,\text{yr}^{-1}$. Their results therefore compare well to ours, as expected, and point towards potentially many detections per year of BHBH mergers produced by the CHE scenario.

In Figure \ref{fig:bhbhdetect} we show how the detection probability $p_k$ of BHBH mergers varies with the chirp mass $M_{\textrm{chirp}}$ and merger redshift $z_m$ at aLIGO's full design sensitivity. Note how quickly the detection probability drops to essentially zero as the redshift increases. Figure \ref{fig:chirpmass} shows the chirp mass distribution of mergers detectable by an aLIGO run at full design sensitivity. It predicts that massive mergers above the PISN gap should be detected. Furthermore, Figure \ref{fig:full_sense_Mchirp} shows how the detection probability varies as a function of merger redshift for systems both above and below the PISN gap, while Figure \ref{fig:mchirp_35} and \ref{fig:mchirp_180} show the detection probability as a function of merger redshift for the O1, O3 and full aLIGO sensitivities for one system below ($M_\textrm{chirp} = 35 \text{M}_{\odot}$) and one system above ($M_\textrm{chirp} = 180 \text{M}_{\odot}$) the PISN gap, respectively. These three figures in combination with Figure \ref{fig:bhbhdetect} give an idea of the detection probability values of systems with different chirp masses.

\begin{figure*}
    \centering
    \begin{subfigure}[b]{0.47\textwidth}
        \includegraphics[width=\textwidth]{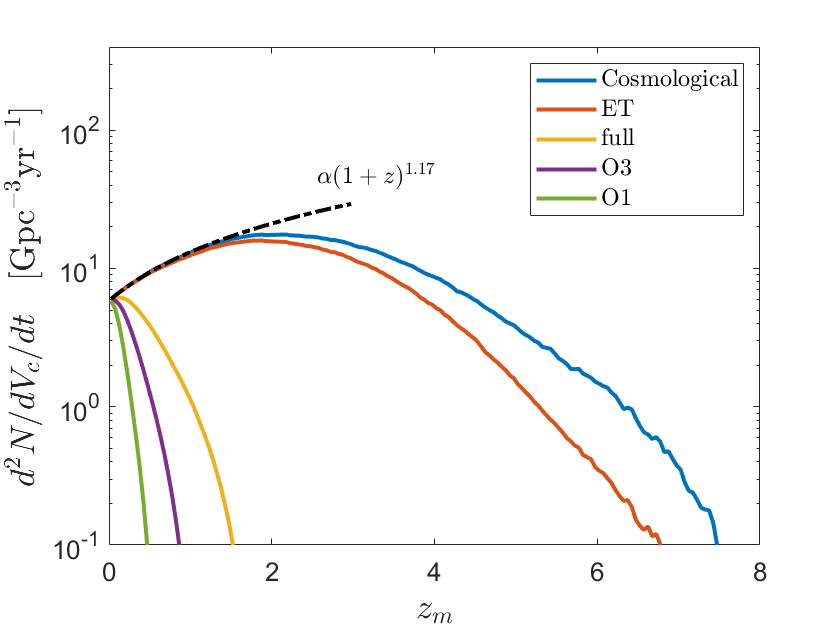}
        \caption{}
        \label{fig:Gpc_rates}
    \end{subfigure}
    ~
    \begin{subfigure}[b]{0.47\textwidth}
        \includegraphics[width=\textwidth]{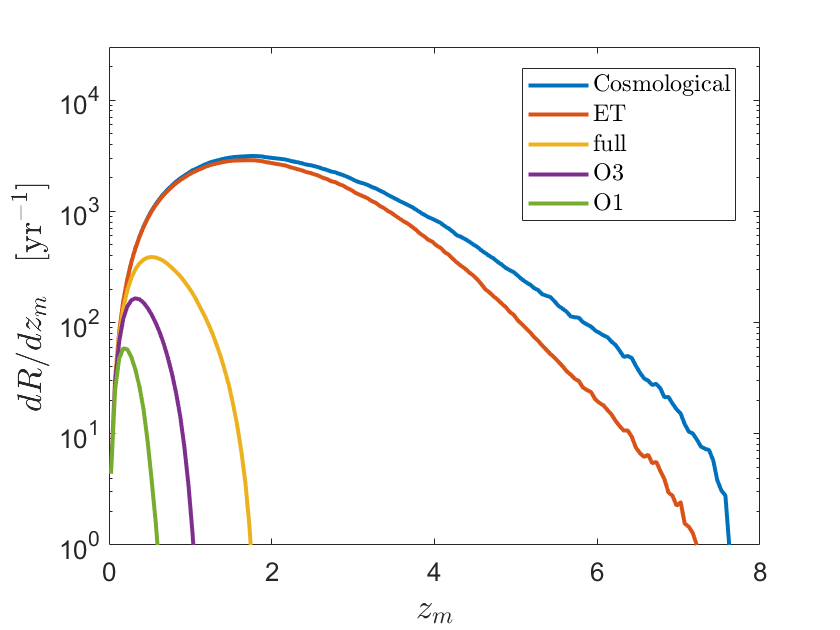}
        \caption{}
        \label{fig:rates}
    \end{subfigure}
    
    \begin{subfigure}[b]{0.47\textwidth}
        \includegraphics[width=\textwidth]{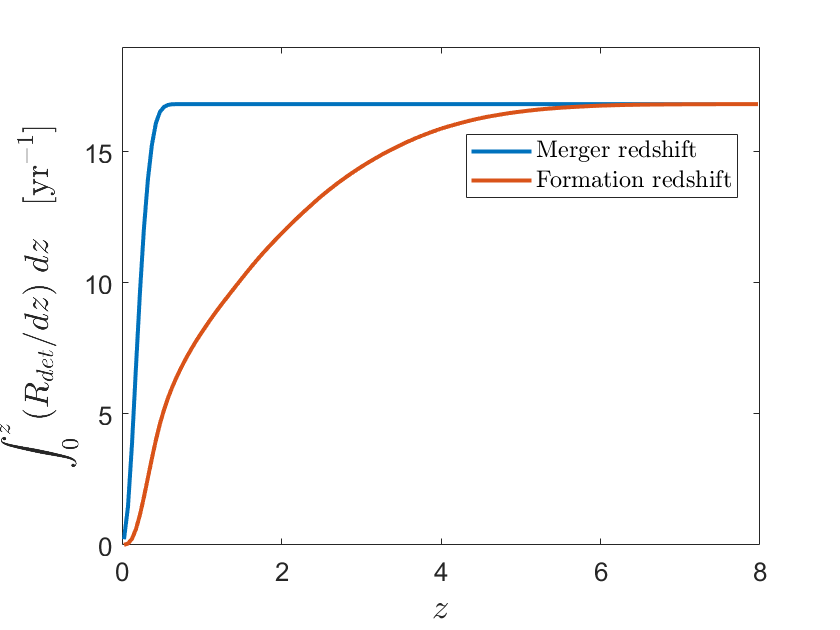}
        \caption{O1 Sensitivity}
        \label{fig:cum_O1}
    \end{subfigure}
    ~
    \begin{subfigure}[b]{0.47\textwidth}
        \includegraphics[width=\textwidth]{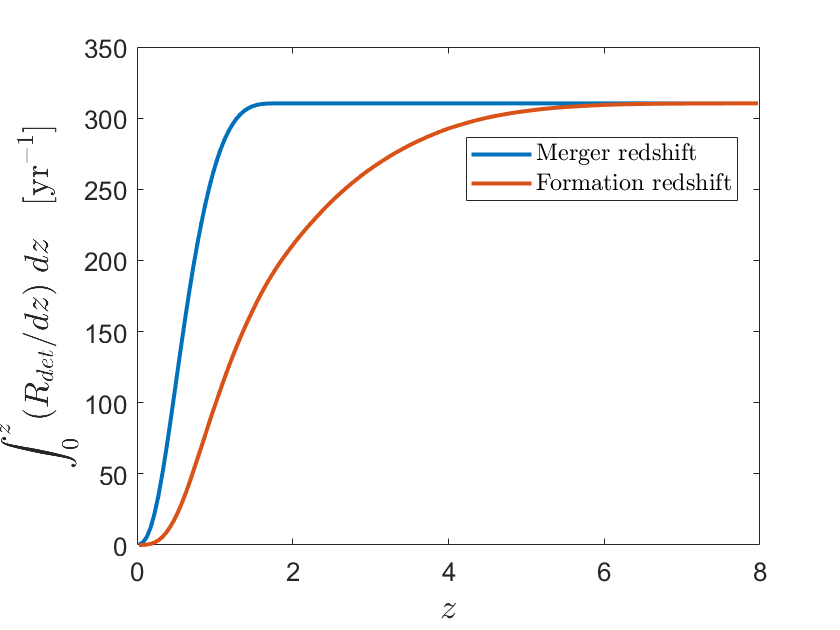}
        \caption{Full sensitivity}
        \label{fig:cum_full}
    \end{subfigure}
    \caption{\textbf{(a)} The cosmological, ET sensitivity and O1-, O3-, and full sensitivity aLIGO co-moving volumetric BHBH merger rates as a function of merger redshift $z_m$. Both the cosmological and detected volumetric rates are here denoted by $d^2 N/dV_c/dt$ for simplicity. Also shown is a power law fit to the cosmological BHBH volumetric merger rate close to $z=0$, for purposes of comparison with the redshift evolution of the SFR close to $z=0$ \citep{madau2014}. \textbf{(b)} The cosmological and detected merger rates observable from Earth (both groups denoted by $R$ for simplicity, see also Eq. \ref{eq:rm_main} and \ref{eq:rm2_main}) binned in terms of merging redshift $z_m$. \textbf{(c)} The cumulative distribution of the total number of O1-detectable mergers observable from Earth per year as a function of both merging and formation redshift. \textbf{(d)} Similar to (c), but for the full aLIGO sensitivity.}
    \label{fig:groupplot2}
\end{figure*}

Figure \ref{fig:Gpc_rates} shows the cosmological, ET sensitivity and \mbox{O1-}, O3-, and full sensitivity aLIGO co-moving volumetric BHBH merger rates (denoted by $d^{2}N/dV_c/dt$) as a function of merger redshift (see Eq. \ref{eq:com_rate} and \ref{eq:aligo_com_rate}), while Figure \ref{fig:rates} shows these rates integrated over our light cone, denoted by $R$ (see Eq. \ref{eq:rm_main}/\ref{eq:rm2_main}). Figure \ref{fig:Gpc_rates} shows that the aLIGO volumetric merger rates decline rapidly with redshift (as suggested by Figures \ref{fig:bhbhdetect} and \ref{fig:groupplot1}) and further that the cosmological co-moving volumetric merger rate peaks around $z \sim 2.5$. We find a local co-moving merger rate of $5.8 \hspace{1mm} \textrm{Gpc}^{-3} \textrm{yr}^{-1}$, to be compared with the aLIGO and Virgo Collaboration's estimated local co-moving volumetric merger rate of ${53.2}_{-28.2}^{+55.8} \hspace{1mm} \text{Gpc}^{-3}\text{yr}^{-1}$ after completion of the O2 run \citep{review2019, a2}. It is also worthwhile investigating whether this BHBH merger rate follows the redshift evolution of the SFR at $z=0$. From \cite{madau2014} we have that the co-moving volumetric rate of star formation follows the relation $\textrm{SFR} \propto (1+z)^{2.7}$ near $z=0$. Similarly assuming a power law form for the cosmological BHBH volumetric merger rate near $z=0$, we find $d^2 N_m/dV_c/dt = 5.8\, (1+z)^{1.17}$. This is illustrated in Figure \ref{fig:Gpc_rates}, thus showing that the BHBH volumetric merger rate evolution differs slightly from that of the SFR.

Figure \ref{fig:rates} shows that the cosmological merger rate observable from Earth peaks around $z \sim 2$ and in Figures \ref{fig:cum_O1} and \ref{fig:cum_full}, the cumulative distribution of the total number of detections per year (detected by the O1 and full design sensitivities, respectively) is shown as a function of both merger and formation redshift of the binary systems, comparing well to that found by \cite{p35}.



\subsection{Momentum kicks}
\label{subsec:kicks}
Up to this point we assumed that massive stars (those not in the PISN regime) either directly collapsed into black holes without any associated loss in mass or energy (direct collapsers) or that they lost mass via multiple pulsations before arriving at their final He-core mass, and collapsed into black holes after this point, again with no losses associated to the actual BH collapse (PPISNe). We could therefore take the final He masses and orbital periods of these binary stars (as shown in Figure \ref{fig:heDepletion}) to represent the masses and periods of the BHBH systems after collapse. In reality, however, the post-collapse parameter values of the resulting BHBH systems also depend on the BH formation process itself. Taking those effects into account will alter the post-collapse parameters, thereby introducing differences in previously calculated merger delay times that might affect eventual BHBH merger population properties.  

BH formation may include a momentum kick similar to those experienced by neutron stars (NSs) at birth \citep{Janka2012}. This is particularly true if the progenitor collapses to a black hole via a two-step process that includes both an explosion and a subsequent collapse by fallback \citep{Brandt1995, Chan2018}. In the case of NS formation, the magnitude of the imparted kicks are fairly well constrained from observations of young radio pulsars \citep{Hobbs2005, igo2018}. This is not the case for BH formation, however: measured masses of observed stellar black holes are relatively small, ranging from $\sim 4 - 50 M_\odot$ \citep{Mc2014,Orosz2007,Prest2007,review2019}, with inferred BH kicks ranging anywhere from virtually nothing \citep{Nelemans1999} to kicks of several $100\, \text{km/s}$ (\citealt{Janka2013, mandelest, repetto1}). For the more massive stellar black holes, like those of particular interest to us, the situation could be quite different since the progenitor stars are more likely to collapse directly into black holes (see \citealt{Fryer1999}) without an accompanying SN explosion, leading to very low kick velocities. Other uncertainties include the amount of mass lost during the collapse, which has a further effect on the post-collapse eccentricity and orbital period of the BHBH system.

To illustratively consider the possible impact of these effects, we administer a kick to all binary systems (not in the PISN regime) upon BH formation using the equations set forth in \cite{Brandt1994}: we assume a uniform distribution in fractional mass loss between $0.0$ and $0.2$, an imparted kick velocity selected from a uniform distribution in the range $0.0 < |v_\textrm{kick}| < 300$ km/s, and a random kick velocity direction. For simplicity, we include only one kick per system (cf. \cite{p36}), but results can easily be generalised. The new post-collapse orbital period, mass and eccentricity of the BHBH systems can then be used to calculate the time delay $t_\textrm{d}$ as shown in \cite{Peters1964}:

\begin{equation}
t_\textrm{d} = \frac{12}{19} \frac{c_0^4}{\beta} \bigintss\limits_0^{e_0} \frac{e^{29/19} [1 + (121/304)e^2]^{1181/2299}}{(1-e^2)^{3/2}} de .
\label{eq:time_delay_kick}
\end{equation}

\noindent In Equation \ref{eq:time_delay_kick}, $e$ is the eccentricity of the system, with $e_0$ being the eccentricity immediately after the collapse; $\beta$ is defined as

\begin{equation}
\beta = \frac{64}{5} \hspace{0.5mm} \frac{G^3 M_\textrm{b1,f} M_\textrm{b2,f} (M_\textrm{b1,f} + M_\textrm{b2,f})}{c^5} ,
\label{eq:beta}
\end{equation}

\noindent with $G$ being the gravitational constant, $M_\textrm{b1,f}$ and $M_\textrm{b2,f}$ the two post-collapse BH masses of the system and $c$ the speed of light. Furthermore, $c_0$ can be calculated using

\begin{equation}
a_0 = \frac{c_0 e_0^{12/19}}{(1-e_0^2)} \hspace{1mm} \left[1 + \frac{121}{304} e_0^2\right]^{870/2299} ,
\label{eq:c_naught}
\end{equation}

\noindent where $e_0$ and $a_0$ are the eccentricity and semi-major axis, respectively, of the orbit after collapse.

It is found that only a very small percentage of BHBH systems formerly produced in our co-moving box by our default Monte Carlo simulation, $6.3 \%$, are completely disrupted by an applied kick, implying that momentum kicks generally have very little effect on the survival rate of these systems. This is expected, as the kicks considered are typically small compared to the orbital velocity of the progenitors. We are, of course, very interested in the effect on eventual BHBH merger populations, as momentum kicks can either lengthen or shorten the post-collapse orbit and so influence the delay time of merger systems. Our results show that, in our co-moving simulation box, $97 \%$ of surviving systems merge on a longer timescale than before, with the rest merging on a shorter timescale. Furthermore, of the systems that originally merged in the lifetime of the Universe, $26 \%$ continue to merge irrespectively, $73 \%$ now merge outside of that timescale, and only $1 \%$ are disrupted completely. In totality, whereas for the original case $45 \%$ of all BHBH systems in our simulations would merge in the lifetime of the Universe, only roughly $12 \%$ now continue to do so.

\begin{figure}
\centering
\includegraphics[width=0.45\textwidth]{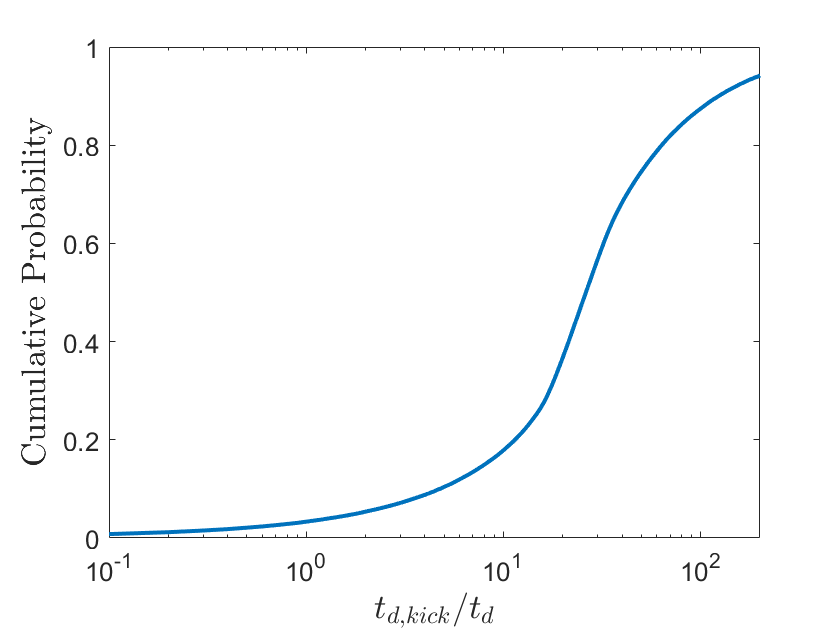}
\caption{The cumulative distribution function of the time delay ratio of BHBH systems in our \textit{PPISN + Kicks} case (mergers and non-mergers alike), where the ratio is taken to be the delay time of a system with an applied kick ($t_{d,kick}$) to the delay time when no kick is applied. Systems with ratios smaller than 1 correspond to binaries whose merging redshift would be higher after a kick is applied (their delay times have been shortened). All other systems will have merger redshifts shifting to lower values, in many cases enough to avoid merging before the current epoch.}
\label{fig:delayRatio}
\end{figure}

Figure \ref{fig:delayRatio} shows the probability distribution of the ratio of BHBH delay times where a kick is applied to that where it is not applied (including all mergers and non-mergers alike) in our simulations. Systems falling below a ratio of 1 will merge at higher redshifts than before, making them more difficult to detect by aLIGO, while other systems will merge at lower redshifts than before. In the latter case, some mergers will shift to redshifts more detectable by aLIGO, while others will shift far enough to not have merged yet (see Figure \ref{fig:bhbhdetect}). In our case, the general trend is for kicks to lengthen the delay time of systems, and so there is an overall downwards shift in the population's merger redshifts. For roughly half of our systems, the delay time is increased by a factor of more than $\sim 20-30$. The general trend of increased delay times means that some systems that merged in the absence of kicks no longer merge within the current age of the Universe. On the other hand, some systems that merged at very high redshifts now merge at lower redshifts, where the mergers could be detectable by aLIGO and ET. These two effects combine to yield a significant decrease in the cosmological merger rate and a less pronounced decrease in the detectable merger rate. Details of this are presented in \mbox{Table \ref{tab:results}.}

\subsection{Pair-instability supernovae}
\label{subsec:PISNe}
As mentioned in Section \ref{sec:intro}, PISNe are rare events that have not been identified observationally in an unambiguous way, although there are various candidate events discussed in the literature \citep{cand2, cand3, cand1, gomez2019}. These events could help us to better understand the evolution of massive stars in low-metallicity regimes. Using the methods and calculations outlined in Section \ref{subsubsec:calcs} and Appendix \ref{subsec:rates}, we calculate the intrinsic PISN rate along with various PISN detection rates in magnitude limited surveys in order to estimate whether these events are likely to be detected by available instruments. We explore peak bolometric magnitude ($m_{\textrm{bol}}^{\textrm{lim}}$) limits of $17, 19, 21, 23$ and $25$, corresponding to a wide range of completed surveys (those with limiting magnitudes of up to roughly $21$, e.g. the Sloan Digital Sky Survey (SDSS)\footnote{SDSS home: \url{http://www.sdss.org}} and the Palomar Transient Factory (PTF)\footnote{PTF home: \url{http://www.ptf.caltech.edu}}), ongoing deeper surveys (that can observe to magnitudes of up to roughly $23$, e.g. the Dark Energy Survey (DES)\footnote{DES home: \url{http://www.darkenergysurvey.org}}, the Panoramic Survey Telescope And Rapid Response System (Pan-STARRS)\footnote{Pan-STARRS home: \url{http://pswww.ifa.hawaii.edu/pswww}} as well as deeper surveys like the Subaru Hyper Suprime-Cam (HSC)\footnote{HSC home: \url{http://hsc.mtk.nao.ac.jp/ssp/}}), and future surveys (that will observe up to magnitudes of roughly $25$, e.g. the Large Synoptic Survey Telescope (LSST)\footnote{LSST home: \url{http://www.lsst.org}} and the Zwicky Transient Facility (ZTF)\footnote{ZTF home: \url{https://www.ztf.caltech.edu}}). These results are given in \mbox{Table \ref{tab:results_pisne},} where both the PISN and CCSN rates are given for the intrinsic and magnitude limited cases.

The numbers in Table \ref{tab:results_pisne} show that the intrinsically fainter PISNe ($M_{\textrm{bol}}^{\textrm{peak}} > -19$) will remain for the most part unobserved. The detection rate of PISNe is dominated by the intrinsically bright events ($M_{\textrm{bol}}^{\textrm{peak}} \leq -19$), which would be observed as superluminous Type I SNe. As bright PISNe originate from the most massive stars, they will be tracing the IMF at the highest stellar masses. A detection of such supernovae at a rate comparable to those given in Table \ref{tab:results_pisne} would strongly support our predicted BH merger rate for BH masses above the PISN gap. \cite{nicholl2013} find that the lack of unambiguous nearby PISN events suggests a local rate of occurrence of less than $6 \times 10^{-6}$ times that of the CCSN rate. This is a factor of 5 smaller than the ratio we find, and could be due to our model assumptions or to PISNe being misclassified in observational data.

\begin{figure}
\centering
\includegraphics[width=0.45\textwidth]{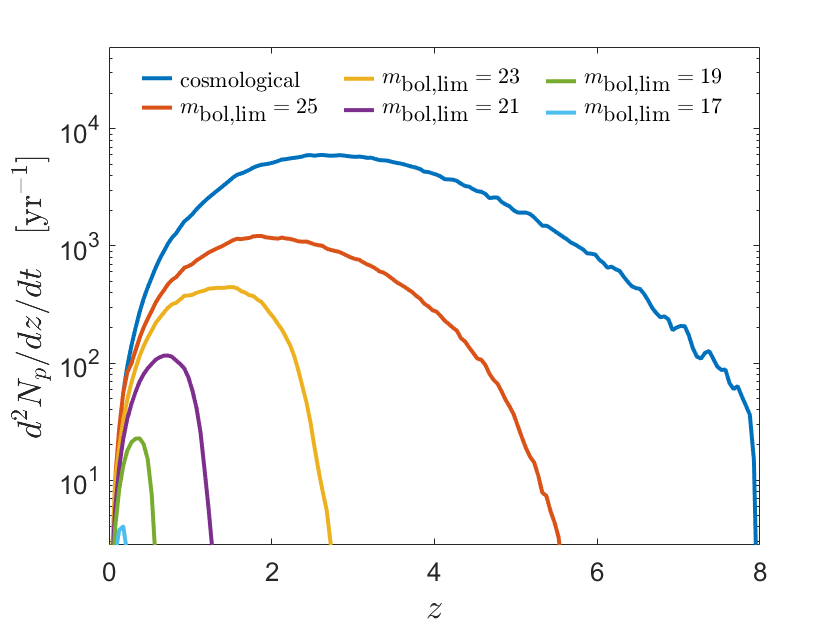}
\caption{PISN rates (for the intrinsic/cosmological and magnitude limited cases) as a function of redshift $z$, as discussed in Section \ref{subsec:PISNe}. There is a clear intrinsic peak at $z \sim 3$, and it can be seen that a significant number of PISNe occur at higher redshifts.} 
\label{fig:PISNrates}
\end{figure}

For CCSNe, we assume a peak bolometric magnitude $M_{\textrm{bol,Ni}}^{\textrm{peak}}$ of $-17.0$ and then follow the same procedure as for PISNe.  \mbox{Figure \ref{fig:PISNrates}} further shows the PISN rates (both for the intrinsic and magnitude limited cases) as a function of redshift, showing a clear intrinsic peak at $z \sim 3$ and a significant number of PISNe occurring at high redshifts. We note that our channel may not be the only one producing PISNe - they could also form from very massive single stars at low metallicity \citep{langer2007,Yusof2013} or via the merger of massive evolved stars \citep{vigna2019}.


\begin{table}
\centering
\caption{PISN and CCSNe rates (for the intrinsic/cosmological and magnitude limited cases) in units of $\textrm{yr}^{-1}$ for the default configuration: bright PISNe have an absolute peak bolometric magnitude $M_{\textrm{bol}}^{\textrm{peak}} \leq -19$, while faint PISNe have $M_{\textrm{bol}}^{\textrm{peak}} > -19$. The last column shows the total number of PISNe expected per $1000$ CCSNe for each of the magnitude limited cases, as well as for the cosmological case. The values in this table are given for the whole sky.}
\label{tab:results_pisne}
    \begin{tabular}{@{}lcccc}
    \hline
    ~        & Bright PISNe & Faint PISNe & CCSNe & CC ratio\\ \hline
    Cosmological & $8070$ & $12000$ & $7.51 \times 10^8$ & $0.027$  \\
    $m_{\textrm{bol}}^{\textrm{lim}} = 25$  & $2970$ & $44.4$ & $2.46 \times 10^6$ & $1.23$  \\
    $m_{\textrm{bol}}^{\textrm{lim}} = 23$  & $643$ & $4.15$ & $2.28 \times 10^5$ & $2.84$ \\
     $m_{\textrm{bol}}^{\textrm{lim}} = 21$  & $79.5$ & $0.335$ & $17500$ & $4.56$ \\
     $m_{\textrm{bol}}^{\textrm{lim}} = 19$  & $7.79$ & $0.0243$ & $1240$ & $6.32$  \\
    $m_{\textrm{bol}}^{\textrm{lim}} = 17$ & $0.659$ & $0.00164$ & $82.5$ & $8.01$ \\ \hline
    \end{tabular}
\end{table}


\subsection{High-redshift deviations in SFR}
\label{subsec:high_redshift}
Finally, we investigate the effects of deviations in the SFR at high redshift on the BHBH merger population from the CHE channel.  Although lower-redshift SFR uncertainties and the uncertainty in the metallicity-specific star formation history could also have a significant impact on merger rates \citep{Neijssel2019,Chruslinska2019a}, we choose to explore here the effects of deviations in the high-redshift SFR only.  The SFR at high redshifts is presently poorly constrained by observations.  In order to investigate the impact of high-redshift SFR variations on the merger rate prediction, we modulate the SFR by multiplying our default model by a sigmoid function.  The sigmoid multiplier is designed to limit changes to the SFR to high redshifts, beyond the peak at $z \sim 2.5$:

\begin{equation}
f(z) = \frac{A}{1 + e^{-az + b}} + 1.
\label{eq:sigmoid}
\end{equation}

\begin{figure}
\centering
\includegraphics[width=0.45\textwidth]{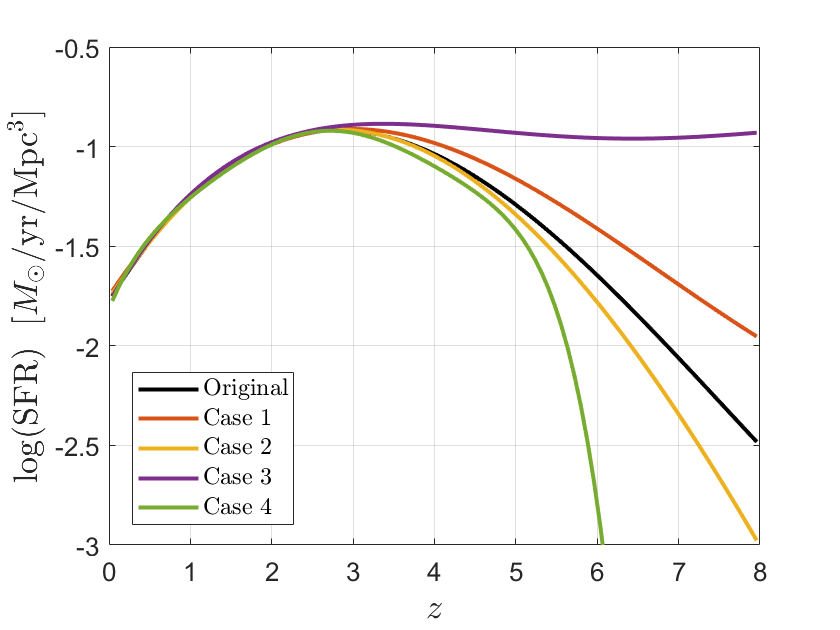}
\caption{The various adopted deviations in high-redshift SFR. To obtain these four cases, the original SFR (represented here by the black curve) was modulated by the sigmoid function given in Equation \ref{eq:sigmoid} by adjusting the value of $A$, $a$ and $b$. \textbf{Case 1:} $A = 2.3306$, $a = 1.25$, $b = 8.2$; the original SFR is increased by up to a factor of 3 at $z = 8$. \textbf{Case 2:} $A = -0.7769$, $a = 1.25$, $b = 8.2$; the original SFR is decreased by up to a factor of $1/3$ at $z = 8$. \textbf{Case 3:} $A = 84.7591$, $a = 1.25$, $b = 10.5$; the SFR remains roughly constant from its peak onwards. \textbf{Case 4:} $A = -5.2990$, $a = 1.25$, $b = 9.0$; the SFR quickly drops to zero after its peak.}
\label{fig:SFRdev}
\end{figure}

\noindent We explore four scenarios: Case 1 gradually increases the original SFR up to a factor of 3 at $z = 8$; Case 2 gradually decreases the original SFR down to a factor of 1/3 at $z = 8$; Case 3 assumes a roughly constant SFR from the peak onwards; Case 4 assumes that the SFR quickly drops to zero after its peak. Figure \ref{fig:SFRdev} shows each of these four cases. The global CCSN rate is adjusted for each case.  We assume that the metallicity distribution is unaffected by changes to the SFR.


\begin{table} \centering
\caption{The cosmological, full aLIGO and ET design sensitivity BHBH merger rate results (in units of $\textrm{yr}^{-1}$) for SFR deviations at high redshifts (see \mbox{Figure \ref{fig:SFRdev}} for the four different SFR cases). Total merger rates are further broken down into mergers above and below the PISN gap. The simulation results for the original SFR are also given for easy comparison (see Table \ref{tab:results}).}
\label{tab:results_sfr}
    \begin{tabular}{@{}lcccc}
    \hline
    ~        & ~              & Cosmological & Full aLIGO & ET \\ \hline
    ~        & Total          & 8510  & 310 & 7130 \\
    Original & Above PISN gap & 1830 & 65.0 & 1220  \\
    ~        & Below PISN gap & 6680 & 245 & 5910  \\ \hline
 ~        & Total             & 9190  & 311 & 7540 \\
    Case 1 & Above PISN gap   & 2020 & 64.5 & 1250 \\
    ~        & Below PISN gap & 7170 & 246 & 6290 \\ \hline
    ~        & Total          & 8270 & 310 & 6980 \\
    Case 2  & Above PISN gap  & 1760 & 65.0 & 1210   \\
    ~        & Below PISN gap & 6510 & 245 & 5770 \\ \hline
    ~        & Total          & 11900 & 314 & 9110   \\
    Case 3  & Above PISN gap  & 2840 & 63.3 & 1360  \\
    ~        & Below PISN gap & 9070 & 251 & 7750  \\ \hline
    ~        & Total          & 7780 & 309 & 6670  \\
    Case 4  & Above PISN gap  & 1640 & 65.4 & 1190  \\
    ~        & Below PISN gap & 6140 & 244 & 5480   \\ \hline
    \end{tabular}
\end{table}

\begin{figure}
\centering
\includegraphics[width=0.45\textwidth]{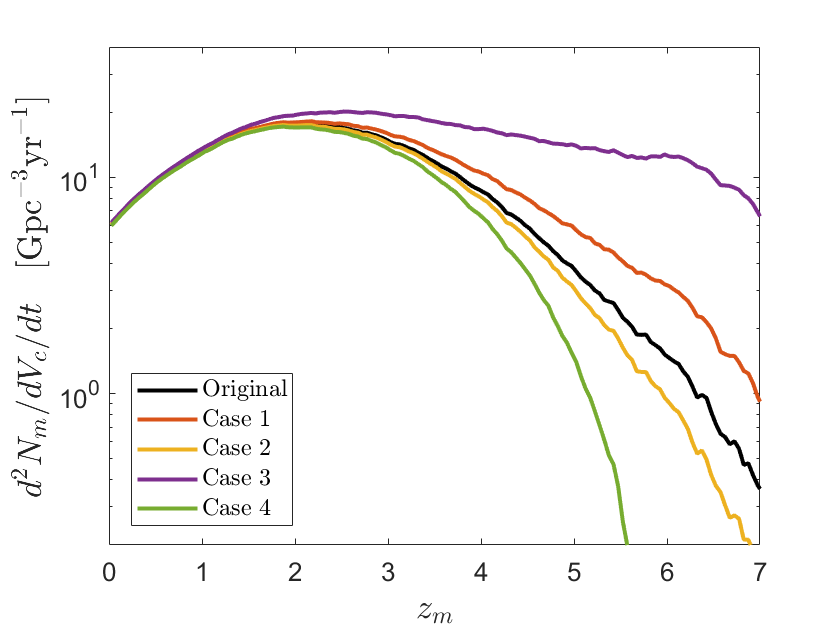}
\caption{The co-moving cosmological BHBH merger rate for the default SFR case (labeled ``Original" in the figure) as well as for each of the four cases of high-redshift deviations in SFR, as a function of merger redshift $z_m$. To see the four cases, see \mbox{Figure \ref{fig:SFRdev}.}}
\label{fig:cases}
\end{figure}


Results are shown in Figure \ref{fig:cases} and \mbox{Table \ref{tab:results_sfr}}. The overall higher SFRs of Cases 1 and 3 increase the total (cosmological) BHBH merger rates, while the opposite is true of Cases 2 and 4. As expected, Cases 3 and 4 result in the most extreme deviations from the original BHBH merger rate. For aLIGO's full sensitivity, the rate deviations from the original model are very small; this is due to the fact that SFR deviations at high redshifts influence the formation of binary stars only at high redshifts, which then have to merge between a redshift of \mbox{0 and $\sim 1.5$} to have a chance of being detected by aLIGO (see Figure \ref{fig:full_sense_Mchirp}). Figure \ref{fig:cases} shows that the cosmological merger rate is essentially unchanged below $z_m = 1.5$, explaining the lack of changes in the population detectable by aLIGO. For all cases but Case 3, the cosmological BHBH merger rate variation is less than or around $\sim 8$\%; for Case 3 however, the variation is more prominent with a $\sim 40\%$ change.

\section{Concluding remarks}
\label{sec:conclusion}
We have performed detailed Monte Carlo simulations in order to investigate the population properties, rate estimates and aLIGO detection rates of BHBH mergers evolving through the CHE channel in the version proposed by \cite{p36}. The results from their detailed simulations, which we vastly extended for this study, as well as the SFR/metallicity distributions from the cosmological simulations of \citet{2015MNRAS.448.1835T}, were used for this purpose.

The cosmological and ET sensitivity BHBH merger rates as well as the O1, O3, and full design sensitivity aLIGO detection rates obtained are summarized in Table \ref{tab:results}. At full aLIGO sensitivity, \cite{p36}'s predicted BHBH detection rate ranges of $2.1 - 370\,\text{yr}^{-1}$ above and $19 - 550\,\text{yr}^{-1}$ below the PISN gap. These compare well to our estimates of $65\,\text{yr}^{-1}$ above and $245\,\text{yr}^{-1}$ below the PISN gap in the case where we include the effects of pulsational-PISNe (PPISNe) and $65\,\text{yr}^{-1}$ above and $317\,\text{yr}^{-1}$ below the PISN gap in the case where we omit the effects of PPISNe, as is expected. Our results are more precise as we investigated a continuous range of metallicities and used the joint SFR and metallicity distributions from the detailed, realistic cosmological simulations of \cite{2015MNRAS.448.1835T}. From Figure \ref{fig:Gpc_rates} we predict a local co-moving merger rate of $5.8\,\textrm{Gpc}^{-3} \textrm{yr}^{-1}$, comparing well to estimates and predictions by \cite{review2019} and \cite{p21}. Our merger rate estimate translates into $\sim 2$ expected detections of BHBH mergers through the CHE channel during aLIGO's O1 observing run, when a total of three BHBH mergers were detected. It is further interesting to note that our results point to the possible detection of massive BHBH mergers from above the PISN gap (see Figure \ref{fig:chirpmass}). At our estimated full sensitivity merger rate of $310\,\text{yr}^{-1}$ for the default case where we include the effects of PPISNe (with a significant number of those coming from above the PISN gap - see Table \ref{tab:results_sfr}), mergers resulting from the CHE scenario could very well turn out to be a significant source of aLIGO detections. As mentioned, we also estimated the detection rates for the projected ET sensitivity and found that the ET would be able to detect $83.8\%$ of BHBH mergers at the present epoch, an astounding improvement on aLIGO.

Our simulation results further give an insight as to the population properties of BHBH systems eventually merging at the present epoch. Figure \ref{fig:heDepletion} shows that the range of final orbital periods for systems at high metallicity is larger than for those at low metallicities, a direct consequence of the strong metallicity dependence of stellar wind mass loss. It further shows that systems with delay times short enough to merge within the lifetime of the Universe have predominantly low metallicities (less than $\sim Z_\odot/10$). At very low metallicities, systems can have very short delay times (see \mbox{Figure \ref{fig:delay_times}}), pointing to the possibility of future high-redshift detections and an opportunity to probe the evolution of massive stars in the early Universe.

Furthermore, natural by-products and variants of our Monte Carlo code enabled us to calculate the effects of momentum kicks during black-hole formation on the population properties of BHBH progenitors, cosmological and magnitude limited PISN rates, and the effects of deviations in the SFR at high redshifts. It is found that the overall effect of momentum kicks is such that it tends to increase the delay time of a BHBH system, as is shown in \mbox{Figure \ref{fig:delayRatio}}. This causes a general downwards shift in the population's merger redshifts (moving some of the systems into the redshift ranges detectable by aLIGO and ET) meaning that the observed percentage decrease in aLIGO and ET merger detections is less pronounced than the accompanying decrease in cosmological mergers. We furthermore calculated various magnitude limited PISN detection rates, which show that with current ongoing deeper surveys and future surveys like the LSST, these rare events should be detectable. Lastly, in studying high-redshift deviations in the SFR and the effects it has on BHBH merger rates, it is found that in cases of mild SFR deviations the effects on the overall cosmological BHBH merger rate is not significant (merger rate estimates change only at the $\sim 8\%$ level), but that it is pronounced in cases of extreme deviation (with changes of up to $\sim 40\%$).

\section*{Acknowledgements}
\label{sec:acknowledgements}
LdB  would like to acknowledge support from the Rhodes Trust and Christ Church. NL's Alexander von Humboldt Professorship and PP's Humboldt Research Award provided essential support for this research. PM acknowledges support from NSF grant AST-1517753 to Vicky Kalogera at Northwestern University, and the FWO junior postdoctoral fellowship No. 12ZY520N. CK acknowledges funding from the UK Science and Technology Facility Council (STFC) through grant ST/M000958/1 \& ST/ R000905/1.
Her work used the DiRAC Data Centric system at Durham University, operated by the Institute for Computational Cosmology on behalf of the STFC DiRAC HPC Facility (www.dirac.ac.uk). This equipment was funded by a BIS National E-infrastructure capital grant ST/K00042X/1, STFC capital grant ST/ K00087X/1, DiRAC Operations grant ST/K003267/1 and Durham University. DiRAC is part of the National E-Infrastructure. IM was supported in part by the STFC and is a recipient of Australian Research Council Future Fellowship FT190100574 funded by the Australian government. SdM acknowledges funding by the European Union's Horizon 2020 research and innovation program from the European Research Council (ERC, grant agreement No.\ 715063), and by the Netherlands Organisation for Scientific Research (NWO) as part of the Vidi research program BinWaves with project number 639.042.728. SdM also acknowledges M. Renzo and F. Broekgaarden for helpful discussions. TJM is supported by the Grants-in-Aid for Scientific Research of the Japan Society for the Promotion of Science (JP17H02864, JP18K13585). Parts of this research were supported by the Australian Research Council Centre of Excellence for All Sky Astrophysics in 3 Dimensions (ASTRO 3D), through project number CE170100013. The Monte Carlo simulation was run on the University College London's high-performance computing facility associated with the Astrophysics department.

\section*{Data availability}
\label{sec:data_avail}
The data underlying this article are available in Zenodo, a general-purpose open-access repository, at \url{https://doi.org/10.5281/zenodo.3348337}.

\bibliographystyle{mnras}
\bibliography{ms}

\begin{thebibliography}{}
\makeatletter
\relax
\def\mn@urlcharsother{\let\do\@makeother \do\$\do\&\do\#\do\^\do\_\do\%\do\~}
\def\mn@doi{\begingroup\mn@urlcharsother \@ifnextchar [ {\mn@doi@}
  {\mn@doi@[]}}
\def\mn@doi@[#1]#2{\def\@tempa{#1}\ifx\@tempa\@empty \href
  {http://dx.doi.org/#2} {doi:#2}\else \href {http://dx.doi.org/#2} {#1}\fi
  \endgroup}
\def\mn@eprint#1#2{\mn@eprint@#1:#2::\@nil}
\def\mn@eprint@arXiv#1{\href {http://arxiv.org/abs/#1} {{\tt arXiv:#1}}}
\def\mn@eprint@dblp#1{\href {http://dblp.uni-trier.de/rec/bibtex/#1.xml}
  {dblp:#1}}
\def\mn@eprint@#1:#2:#3:#4\@nil{\def\@tempa {#1}\def\@tempb {#2}\def\@tempc
  {#3}\ifx \@tempc \@empty \let \@tempc \@tempb \let \@tempb \@tempa \fi \ifx
  \@tempb \@empty \def\@tempb {arXiv}\fi \@ifundefined
  {mn@eprint@\@tempb}{\@tempb:\@tempc}{\expandafter \expandafter \csname
  mn@eprint@\@tempb\endcsname \expandafter{\@tempc}}}

\bibitem[\protect\citeauthoryear{{Abadie} et~al.,}{{Abadie} et~al.}{2010}]{p12}
{Abadie} J.,  et~al., 2010, \mn@doi [Classical and Quantum Gravity]
  {10.1088/0264-9381/27/17/173001}, \href
  {http://adsabs.harvard.edu/abs/2010CQGra..27q3001A} {27, 173001}

\bibitem[\protect\citeauthoryear{{Abbott} et~al.,}{{Abbott}
  et~al.}{2016a}]{aLIGOcurves2016}
{Abbott} B.~P.,  et~al., 2016a, \mn@doi [Living Reviews in Relativity]
  {10.1007/lrr-2016-1}, \href
  {http://adsabs.harvard.edu/abs/2016LRR....19....1A} {19, 1}

\bibitem[\protect\citeauthoryear{{Abbott} et~al.,}{{Abbott}
  et~al.}{2016b}]{p21}
{Abbott} B.~P.,  et~al., 2016b, \mn@doi [ApJL] {10.3847/2041-8205/818/2/L22},
  \href {http://adsabs.harvard.edu/abs/2016ApJ...818L..22A} {818, L22}

\bibitem[\protect\citeauthoryear{{Abbott}, {Abbott}, {Abbott}, {Abernathy},
  {Ackley}  et~al.}{{Abbott} et~al.}{2017a}]{CE}
{Abbott} B.~P.,  {Abbott} R.,  {Abbott} T.~D.,  {Abernathy} M.~R.,  {Ackley}
  K.,   et~al., 2017a, \mn@doi [Classical and Quantum Gravity]
  {10.1088/1361-6382/aa51f4}, \href
  {https://ui.adsabs.harvard.edu/abs/2017CQGra..34d4001A} {34, 044001}

\bibitem[\protect\citeauthoryear{{Abbott} et~al.,}{{Abbott}
  et~al.}{2017b}]{CosmicExplorer}
{Abbott} B.~P.,  et~al., 2017b, \mn@doi [Classical and Quantum Gravity]
  {10.1088/1361-6382/aa51f4}, \href
  {https://ui.adsabs.harvard.edu/abs/2017CQGra..34d4001A} {34, 044001}

\bibitem[\protect\citeauthoryear{{Abbott} et~al.,}{{Abbott}
  et~al.}{2017c}]{smallmerge2017}
{Abbott} B.~P.,  et~al., 2017c, \mn@doi [\apjl] {10.3847/2041-8213/aa9f0c},
  \href {http://adsabs.harvard.edu/abs/2017ApJ...851L..35A} {851, L35}

\bibitem[\protect\citeauthoryear{{Abbott} et~al.,}{{Abbott}
  et~al.}{2019a}]{abbott2019}
{Abbott} R.,  et~al., 2019a, arXiv e-prints, \href
  {https://ui.adsabs.harvard.edu/abs/2019arXiv191211716T} {p. arXiv:1912.11716}

\bibitem[\protect\citeauthoryear{Abbott et~al.,}{Abbott
  et~al.}{2019b}]{review2019}
Abbott B.~P.,  et~al., 2019b, \mn@doi [Phys. Rev. X]
  {10.1103/PhysRevX.9.031040}, 9, 031040

\bibitem[\protect\citeauthoryear{Abbott et~al.,}{Abbott et~al.}{2019c}]{a2}
Abbott B.~P.,  et~al., 2019c, \mn@doi [The Astrophysical Journal]
  {10.3847/2041-8213/ab3800}, 882, L24

\bibitem[\protect\citeauthoryear{{Abbott}, {Abbott}, {Abraham}, {Acernese}
  et~al.}{{Abbott} et~al.}{2020a}]{GW190521}
{Abbott} R.,  {Abbott} T.~D.,  {Abraham} S.,  {Acernese} F.,   et~al., 2020a,
  arXiv e-prints, \href {https://ui.adsabs.harvard.edu/abs/2020arXiv200901075T}
  {p. arXiv:2009.01075}

\bibitem[\protect\citeauthoryear{{Abbott}, {Abbott}, {Abraham}, {Acernese},
  {Ackley}  et~al.}{{Abbott} et~al.}{2020b}]{GW190521:astro}
{Abbott} R.,  {Abbott} T.~D.,  {Abraham} S.,  {Acernese} F.,  {Ackley} K.,
  et~al., 2020b, arXiv e-prints, \href
  {https://ui.adsabs.harvard.edu/abs/2020arXiv200901190T} {p. arXiv:2009.01190}

\bibitem[\protect\citeauthoryear{{Ajith} et~al.,}{{Ajith}
  et~al.}{2011}]{Ajith2011}
{Ajith} P.,  et~al., 2011, \mn@doi [Physical Review Letters]
  {10.1103/PhysRevLett.106.241101}, \href
  {http://adsabs.harvard.edu/abs/2011PhRvL.106x1101A} {106, 241101}

\bibitem[\protect\citeauthoryear{{Antonini}, {Chatterjee}, {Rodriguez},
  {Morscher}, {Pattabiraman}, {Kalogera}  \& {Rasio}}{{Antonini}
  et~al.}{2016}]{p33}
{Antonini} F.,  {Chatterjee} S.,  {Rodriguez} C.~L.,  {Morscher} M.,
  {Pattabiraman} B.,  {Kalogera} V.,   {Rasio} F.~A.,  2016, \mn@doi [ApJ]
  {10.3847/0004-637X/816/2/65}, \href
  {http://adsabs.harvard.edu/abs/2016ApJ...816...65A} {816, 65}

\bibitem[\protect\citeauthoryear{{Arnett}}{{Arnett}}{1982}]{Arnett1982}
{Arnett} W.~D.,  1982, \mn@doi [ApJ] {10.1086/159681}, \href
  {http://adsabs.harvard.edu/abs/1982ApJ...253..785A} {253, 785}

\bibitem[\protect\citeauthoryear{{Belczynski}, {Dominik}, {Bulik},
  {O'Shaughnessy}, {Fryer}  \& {Holz}}{{Belczynski} et~al.}{2010}]{p15}
{Belczynski} K.,  {Dominik} M.,  {Bulik} T.,  {O'Shaughnessy} R.,  {Fryer} C.,
   {Holz} D.~E.,  2010, \mn@doi [ApJL] {10.1088/2041-8205/715/2/L138}, \href
  {http://adsabs.harvard.edu/abs/2010ApJ...715L.138B} {715, L138}

\bibitem[\protect\citeauthoryear{{Belczynski}, {Holz}, {Bulik}  \&
  {O'Shaughnessy}}{{Belczynski} et~al.}{2016}]{p27}
{Belczynski} K.,  {Holz} D.~E.,  {Bulik} T.,   {O'Shaughnessy} R.,  2016,
  \mn@doi [Nat] {10.1038/nature18322}, \href
  {http://adsabs.harvard.edu/abs/2016Natur.534..512B} {534, 512}

\bibitem[\protect\citeauthoryear{{Belczynski} et~al.,}{{Belczynski}
  et~al.}{2017}]{Belczynski2017}
{Belczynski} K.,  et~al., 2017, arXiv e-prints, \href
  {https://ui.adsabs.harvard.edu/abs/2017arXiv170607053B} {p. arXiv:1706.07053}

\bibitem[\protect\citeauthoryear{{Blaauw}}{{Blaauw}}{1961}]{Blaauw1961}
{Blaauw} A.,  1961, \bain, \href
  {https://ui.adsabs.harvard.edu/abs/1961BAN....15..265B} {15, 265}

\bibitem[\protect\citeauthoryear{{Brandt} \& {Podsiadlowski}}{{Brandt} \&
  {Podsiadlowski}}{1995}]{Brandt1994}
{Brandt} N.,  {Podsiadlowski} P.,  1995, \mn@doi [MNRAS]
  {10.1093/mnras/274.2.461}, \href
  {http://adsabs.harvard.edu/abs/1995MNRAS.274..461B} {274, 461}

\bibitem[\protect\citeauthoryear{{Brandt}, {Podsiadlowski}  \&
  {Sigurdsson}}{{Brandt} et~al.}{1995}]{Brandt1995}
{Brandt} W.~N.,  {Podsiadlowski} P.,   {Sigurdsson} S.,  1995, \mn@doi [MNRAS]
  {10.1093/mnras/277.1.L35}, \href
  {http://adsabs.harvard.edu/abs/1995MNRAS.277L..35B} {277, L35}

\bibitem[\protect\citeauthoryear{{Brott} et~al.,}{{Brott} et~al.}{2011a}]{p43}
{Brott} I.,  et~al., 2011a, \mn@doi [A\&A] {10.1051/0004-6361/201016113}, \href
  {http://ukads.nottingham.ac.uk/abs/2011A%26A...530A.115B} {530, A115}

\bibitem[\protect\citeauthoryear{{Brott} et~al.,}{{Brott}
  et~al.}{2011b}]{brott2011}
{Brott} I.,  et~al., 2011b, \mn@doi [\aap] {10.1051/0004-6361/201016114}, \href
  {http://adsabs.harvard.edu/abs/2011A%26A...530A.116B} {530, A116}

\bibitem[\protect\citeauthoryear{{Bulik} \& {Belczy{\'n}ski}}{{Bulik} \&
  {Belczy{\'n}ski}}{2003}]{p17}
{Bulik} T.,  {Belczy{\'n}ski} K.,  2003, \mn@doi [ApJL] {10.1086/375713}, \href
  {http://adsabs.harvard.edu/abs/2003ApJ...589L..37B} {589, L37}

\bibitem[\protect\citeauthoryear{{Burgay} et~al.,}{{Burgay} et~al.}{2003}]{p7}
{Burgay} M.,  et~al., 2003, \mn@doi [Nat] {10.1038/nature02124}, \href
  {http://adsabs.harvard.edu/abs/2003Natur.426..531B} {426, 531}

\bibitem[\protect\citeauthoryear{{Casagrande}, {Portinari}  \&
  {Flynn}}{{Casagrande} et~al.}{2006}]{Casagrande2006}
{Casagrande} L.,  {Portinari} L.,   {Flynn} C.,  2006, \mn@doi [\mnras]
  {10.1111/j.1365-2966.2006.10999.x}, \href
  {https://ui.adsabs.harvard.edu/abs/2006MNRAS.373...13C} {373, 13}

\bibitem[\protect\citeauthoryear{{Chan}, {M{\"u}ller}, {Heger}, {Pakmor}  \&
  {Springel}}{{Chan} et~al.}{2018}]{Chan2018}
{Chan} C.,  {M{\"u}ller} B.,  {Heger} A.,  {Pakmor} R.,   {Springel} V.,  2018,
  \mn@doi [\apjl] {10.3847/2041-8213/aaa28c}, \href
  {https://ui.adsabs.harvard.edu/abs/2018ApJ...852L..19C} {852, L19}

\bibitem[\protect\citeauthoryear{{Chruslinska} \& {Nelemans}}{{Chruslinska} \&
  {Nelemans}}{2019}]{Chruslinska2019b}
{Chruslinska} M.,  {Nelemans} G.,  2019, \mn@doi [\mnras]
  {10.1093/mnras/stz2057}, \href
  {https://ui.adsabs.harvard.edu/abs/2019MNRAS.488.5300C} {488, 5300}

\bibitem[\protect\citeauthoryear{{Chruslinska}, {Nelemans}  \&
  {Belczynski}}{{Chruslinska} et~al.}{2019}]{Chruslinska2019a}
{Chruslinska} M.,  {Nelemans} G.,   {Belczynski} K.,  2019, \mn@doi [\mnras]
  {10.1093/mnras/sty3087}, \href
  {https://ui.adsabs.harvard.edu/abs/2019MNRAS.482.5012C} {482, 5012}

\bibitem[\protect\citeauthoryear{{Dominik} et~al.,}{{Dominik}
  et~al.}{2015}]{p16}
{Dominik} M.,  et~al., 2015, \mn@doi [ApJ] {10.1088/0004-637X/806/2/263}, \href
  {http://adsabs.harvard.edu/abs/2015ApJ...806..263D} {806, 263}

\bibitem[\protect\citeauthoryear{{Eddington}}{{Eddington}}{1924}]{eddington1924}
{Eddington} A.~S.,  1924, \mn@doi [\mnras] {10.1093/mnras/84.5.308}, \href
  {https://ui.adsabs.harvard.edu/abs/1924MNRAS..84..308E} {84, 308}

\bibitem[\protect\citeauthoryear{{Farmer}, {Renzo}, {de Mink}, {Marchant}  \&
  {Justham}}{{Farmer} et~al.}{2019}]{Farmer2019}
{Farmer} R.,  {Renzo} M.,  {de Mink} S.~E.,  {Marchant} P.,   {Justham} S.,
  2019, \mn@doi [\apj] {10.3847/1538-4357/ab518b}, \href
  {https://ui.adsabs.harvard.edu/abs/2019ApJ...887...53F} {887, 53}

\bibitem[\protect\citeauthoryear{{Farr}, {Stevenson}, {Miller}, {Mandel},
  {Farr}  \& {Vecchio}}{{Farr} et~al.}{2017}]{Farr}
{Farr} W.~M.,  {Stevenson} S.,  {Miller} M.~C.,  {Mandel} I.,  {Farr} B.,
  {Vecchio} A.,  2017, \mn@doi [Nature] {10.1038/nature23453}, \href
  {http://adsabs.harvard.edu/abs/2017Natur.548..426F} {548, 426}

\bibitem[\protect\citeauthoryear{{Fowler} \& {Hoyle}}{{Fowler} \&
  {Hoyle}}{1964}]{FowlerHoyle1964}
{Fowler} W.~A.,  {Hoyle} F.,  1964, \mn@doi [\apjs] {10.1086/190103}, \href
  {http://adsabs.harvard.edu/abs/1964ApJS....9..201F} {9, 201}

\bibitem[\protect\citeauthoryear{{Fraley}}{{Fraley}}{1968}]{Fraley1968}
{Fraley} G.~S.,  1968, \mn@doi [\apss] {10.1007/BF00651498}, \href
  {http://adsabs.harvard.edu/abs/1968Ap%26SS...2...96F} {2, 96}

\bibitem[\protect\citeauthoryear{Fryer}{Fryer}{1999}]{Fryer1999}
Fryer C.~L.,  1999, \mn@doi [The Astrophysical Journal] {10.1086/307647}, 522,
  413–418

\bibitem[\protect\citeauthoryear{{Gal-Yam} et~al.,}{{Gal-Yam}
  et~al.}{2009}]{cand3}
{Gal-Yam} A.,  et~al., 2009, \mn@doi [\nat] {10.1038/nature08579}, \href
  {https://ui.adsabs.harvard.edu/abs/2009Natur.462..624G} {462, 624}

\bibitem[\protect\citeauthoryear{{Gomez} et~al.,}{{Gomez}
  et~al.}{2019}]{gomez2019}
{Gomez} S.,  et~al., 2019, \mn@doi [\apj] {10.3847/1538-4357/ab2f92}, \href
  {https://ui.adsabs.harvard.edu/abs/2019ApJ...881...87G} {881, 87}

\bibitem[\protect\citeauthoryear{{Grevesse}, {Noels}  \& {Sauval}}{{Grevesse}
  et~al.}{1996}]{in2}
{Grevesse} N.,  {Noels} A.,   {Sauval} A.~J.,  1996, in {Holt} S.~S.,
  {Sonneborn} G.,  eds,  Astronomical Society of the Pacific Conference Series
  Vol. 99, Cosmic Abundances. p.~117

\bibitem[\protect\citeauthoryear{{Heger} \& {Langer}}{{Heger} \&
  {Langer}}{2000}]{p39}
{Heger} A.,  {Langer} N.,  2000, \mn@doi [ApJ] {10.1086/317239}, \href
  {http://ukads.nottingham.ac.uk/abs/2000ApJ...544.1016H} {544, 1016}

\bibitem[\protect\citeauthoryear{{Heger} \& {Woosley}}{{Heger} \&
  {Woosley}}{2002}]{p4}
{Heger} A.,  {Woosley} S.~E.,  2002, \mn@doi [ApJ] {10.1086/338487}, \href
  {http://adsabs.harvard.edu/abs/2002ApJ...567..532H} {567, 532}

\bibitem[\protect\citeauthoryear{{Herzig}, {El Eid}, {Fricke}  \&
  {Langer}}{{Herzig} et~al.}{1990}]{Herzig1990}
{Herzig} K.,  {El Eid} M.~F.,  {Fricke} K.~J.,   {Langer} N.,  1990, AAP, \href
  {http://adsabs.harvard.edu/abs/1990A%26A...233..462H} {233, 462}

\bibitem[\protect\citeauthoryear{{Hinshaw} et~al.,}{{Hinshaw}
  et~al.}{2013}]{hinshaw2013}
{Hinshaw} G.,  et~al., 2013, \mn@doi [\apjs] {10.1088/0067-0049/208/2/19},
  \href {https://ui.adsabs.harvard.edu/abs/2013ApJS..208...19H} {208, 19}

\bibitem[\protect\citeauthoryear{{Hobbs}, {Lorimer}, {Lyne}  \&
  {Kramer}}{{Hobbs} et~al.}{2005}]{Hobbs2005}
{Hobbs} G.,  {Lorimer} D.~R.,  {Lyne} A.~G.,   {Kramer} M.,  2005, \mn@doi
  [MNRAS] {10.1111/j.1365-2966.2005.09087.x}, \href
  {http://adsabs.harvard.edu/abs/2005MNRAS.360..974H} {360, 974}

\bibitem[\protect\citeauthoryear{{Hulse} \& {Taylor}}{{Hulse} \&
  {Taylor}}{1975}]{p5}
{Hulse} R.~A.,  {Taylor} J.~H.,  1975, \mn@doi [ApJL] {10.1086/181708}, \href
  {http://adsabs.harvard.edu/abs/1975ApJ...195L..51H} {195, L51}

\bibitem[\protect\citeauthoryear{{Igoshev} \& {Verbunt}}{{Igoshev} \&
  {Verbunt}}{2018}]{igo2018}
{Igoshev} A.,  {Verbunt} F.,  2018, in American Astronomical Society Meeting
  Abstracts \#231. p. 132.03

\bibitem[\protect\citeauthoryear{{Ivanova} et~al.,}{{Ivanova}
  et~al.}{2013}]{p28}
{Ivanova} N.,  et~al., 2013, \mn@doi [A\&A Rev.] {10.1007/s00159-013-0059-2},
  \href {http://adsabs.harvard.edu/abs/2013A%26ARv..21...59I} {21, 59}

\bibitem[\protect\citeauthoryear{{Janka}}{{Janka}}{2012}]{Janka2012}
{Janka} H.-T.,  2012, \mn@doi [Annual Review of Nuclear and Particle Science]
  {10.1146/annurev-nucl-102711-094901}, \href
  {http://adsabs.harvard.edu/abs/2012ARNPS..62..407J} {62, 407}

\bibitem[\protect\citeauthoryear{{Janka}}{{Janka}}{2013}]{Janka2013}
{Janka} H.~T.,  2013, \mn@doi [MNRAS] {10.1093/mnras/stt1106}, \href
  {http://adsabs.harvard.edu/abs/2013MNRAS.434.1355J} {434, 1355}

\bibitem[\protect\citeauthoryear{{Kalogera}, {Belczynski}, {Kim},
  {O'Shaughnessy}  \& {Willems}}{{Kalogera} et~al.}{2007}]{p26}
{Kalogera} V.,  {Belczynski} K.,  {Kim} C.,  {O'Shaughnessy} R.,   {Willems}
  B.,  2007, \mn@doi [Phys. Rep.] {10.1016/j.physrep.2007.02.008}, \href
  {http://adsabs.harvard.edu/abs/2007PhR...442...75K} {442, 75}

\bibitem[\protect\citeauthoryear{{Kasen}, {Woosley}  \& {Heger}}{{Kasen}
  et~al.}{2011}]{Kasen2011}
{Kasen} D.,  {Woosley} S.~E.,   {Heger} A.,  2011, \mn@doi [ApJ]
  {10.1088/0004-637X/734/2/102}, \href
  {http://adsabs.harvard.edu/abs/2011ApJ...734..102K} {734, 102}

\bibitem[\protect\citeauthoryear{{Kim}, {Kalogera}  \& {Lorimer}}{{Kim}
  et~al.}{2003}]{p10}
{Kim} C.,  {Kalogera} V.,   {Lorimer} D.~R.,  2003, \mn@doi [ApJ]
  {10.1086/345740}, \href {http://adsabs.harvard.edu/abs/2003ApJ...584..985K}
  {584, 985}

\bibitem[\protect\citeauthoryear{{Kobayashi} \& {Nomoto}}{{Kobayashi} \&
  {Nomoto}}{2009}]{2009ApJ...707.1466K}
{Kobayashi} C.,  {Nomoto} K.,  2009, \mn@doi [\apj]
  {10.1088/0004-637X/707/2/1466}, \href
  {https://ui.adsabs.harvard.edu/abs/2009ApJ...707.1466K} {707, 1466}

\bibitem[\protect\citeauthoryear{{Kobayashi}, {Karakas}  \&
  {Umeda}}{{Kobayashi} et~al.}{2011}]{2011MNRAS.414.3231K}
{Kobayashi} C.,  {Karakas} A.~I.,   {Umeda} H.,  2011, \mn@doi [\mnras]
  {10.1111/j.1365-2966.2011.18621.x}, \href
  {https://ui.adsabs.harvard.edu/abs/2011MNRAS.414.3231K} {414, 3231}

\bibitem[\protect\citeauthoryear{{Kobayashi}, {Karakas}  \&
  {Lugaro}}{{Kobayashi} et~al.}{2020}]{2020chiaki}
{Kobayashi} C.,  {Karakas} A.~I.,   {Lugaro} M.,  2020, arXiv e-prints, \href
  {https://ui.adsabs.harvard.edu/abs/2020arXiv200804660K} {p. arXiv:2008.04660}

\bibitem[\protect\citeauthoryear{{K{\"o}hler} et~al.,}{{K{\"o}hler}
  et~al.}{2015}]{p42}
{K{\"o}hler} K.,  et~al., 2015, \mn@doi [A\&A] {10.1051/0004-6361/201424356},
  \href {http://ukads.nottingham.ac.uk/abs/2015A%26A...573A..71K} {573, A71}

\bibitem[\protect\citeauthoryear{{Kroupa}}{{Kroupa}}{2001}]{kroupa2001}
{Kroupa} P.,  2001, \mn@doi [MNRAS] {10.1046/j.1365-8711.2001.04022.x}, \href
  {http://adsabs.harvard.edu/abs/2001MNRAS.322..231K} {322, 231}

\bibitem[\protect\citeauthoryear{{Kulkarni}, {Hut}  \& {McMillan}}{{Kulkarni}
  et~al.}{1993}]{kulka}
{Kulkarni} S.~R.,  {Hut} P.,   {McMillan} S.,  1993, \mn@doi [Nature]
  {10.1038/364421a0}, \href {http://adsabs.harvard.edu/abs/1993Natur.364..421K}
  {364, 421}

\bibitem[\protect\citeauthoryear{{Langer}}{{Langer}}{1992}]{p38}
{Langer} N.,  1992, A\&A, \href
  {http://ukads.nottingham.ac.uk/abs/1992A%26A...265L..17L} {265, L17}

\bibitem[\protect\citeauthoryear{{Langer}, {Norman}, {de Koter}, {Vink},
  {Cantiello}  \& {Yoon}}{{Langer} et~al.}{2007}]{langer2007}
{Langer} N.,  {Norman} C.~A.,  {de Koter} A.,  {Vink} J.~S.,  {Cantiello} M.,
  {Yoon} S.~C.,  2007, \mn@doi [\aap] {10.1051/0004-6361:20078482}, \href
  {https://ui.adsabs.harvard.edu/abs/2007A&A...475L..19L} {475, L19}

\bibitem[\protect\citeauthoryear{{Lipunov}, {Postnov}  \&
  {Prokhorov}}{{Lipunov} et~al.}{1997}]{p13}
{Lipunov} V.~M.,  {Postnov} K.~A.,   {Prokhorov} M.~E.,  1997, \mn@doi [MNRAS]
  {10.1093/mnras/288.1.245}, \href
  {http://adsabs.harvard.edu/abs/1997MNRAS.288..245L} {288, 245}

\bibitem[\protect\citeauthoryear{{Madau} \& {Dickinson}}{{Madau} \&
  {Dickinson}}{2014}]{madau2014}
{Madau} P.,  {Dickinson} M.,  2014, \mn@doi [ARA\&A]
  {10.1146/annurev-astro-081811-125615}, \href
  {http://adsabs.harvard.edu/abs/2014ARA%26A..52..415M} {52, 415}

\bibitem[\protect\citeauthoryear{{Maeder}}{{Maeder}}{1987}]{p37}
{Maeder} A.,  1987, A\&A, \href
  {http://ukads.nottingham.ac.uk/abs/1987A%26A...178..159M} {178, 159}

\bibitem[\protect\citeauthoryear{{Mandel}}{{Mandel}}{2016}]{mandelest}
{Mandel} I.,  2016, \mn@doi [MNRAS] {10.1093/mnras/stv2733}, \href
  {http://adsabs.harvard.edu/abs/2016MNRAS.456..578M} {456, 578}

\bibitem[\protect\citeauthoryear{{Mandel} \& {O'Shaughnessy}}{{Mandel} \&
  {O'Shaughnessy}}{2010}]{p18}
{Mandel} I.,  {O'Shaughnessy} R.,  2010, \mn@doi [Classical and Quantum
  Gravity] {10.1088/0264-9381/27/11/114007}, \href
  {http://adsabs.harvard.edu/abs/2010CQGra..27k4007M} {27, 114007}

\bibitem[\protect\citeauthoryear{{Mandel} \& {de Mink}}{{Mandel} \& {de
  Mink}}{2016}]{p35}
{Mandel} I.,  {de Mink} S.~E.,  2016, \mn@doi [MNRAS] {10.1093/mnras/stw379},
  \href {http://ukads.nottingham.ac.uk/abs/2016MNRAS.458.2634M} {458, 2634}

\bibitem[\protect\citeauthoryear{{Mandel}, {Haster}, {Dominik}  \&
  {Belczynski}}{{Mandel} et~al.}{2015}]{p20}
{Mandel} I.,  {Haster} C.-J.,  {Dominik} M.,   {Belczynski} K.,  2015, \mn@doi
  [MNRAS] {10.1093/mnrasl/slv054}, \href
  {http://adsabs.harvard.edu/abs/2015MNRAS.450L..85M} {450, L85}

\bibitem[\protect\citeauthoryear{{Marchant}, {Langer}, {Podsiadlowski},
  {Tauris}  \& {Moriya}}{{Marchant} et~al.}{2016}]{p36}
{Marchant} P.,  {Langer} N.,  {Podsiadlowski} P.,  {Tauris} T.~M.,   {Moriya}
  T.~J.,  2016, \mn@doi [A\&A] {10.1051/0004-6361/201628133}, \href
  {http://ukads.nottingham.ac.uk/abs/2016A%26A...588A..50M} {588, A50}

\bibitem[\protect\citeauthoryear{{Marchant}, {Langer}, {Podsiadlowski},
  {Tauris}, {de Mink}, {Mandel}  \& {Moriya}}{{Marchant}
  et~al.}{2017}]{Pablo2017}
{Marchant} P.,  {Langer} N.,  {Podsiadlowski} P.,  {Tauris} T.~M.,  {de Mink}
  S.,  {Mandel} I.,   {Moriya} T.~J.,  2017, \mn@doi [A\&A]
  {10.1051/0004-6361/201630188}, \href
  {http://adsabs.harvard.edu/abs/2017A%26A...604A..55M} {604, A55}

\bibitem[\protect\citeauthoryear{{Marchant}, {Renzo}, {Farmer}, {Pappas},
  {Taam}, {de Mink}  \& {Kalogera}}{{Marchant} et~al.}{2018}]{marchant2018}
{Marchant} P.,  {Renzo} M.,  {Farmer} R.,  {Pappas} K. M.~W.,  {Taam} R.~E.,
  {de Mink} S.,   {Kalogera} V.,  2018, arXiv e-prints, \href
  {https://ui.adsabs.harvard.edu/abs/2018arXiv181013412M} {p. arXiv:1810.13412}

\bibitem[\protect\citeauthoryear{{Marchant}, {Renzo}, {Farmer}, {Pappas},
  {Taam}, {de Mink}  \& {Kalogera}}{{Marchant} et~al.}{2019}]{MarchantPPISN}
{Marchant} P.,  {Renzo} M.,  {Farmer} R.,  {Pappas} K. M.~W.,  {Taam} R.~E.,
  {de Mink} S.~E.,   {Kalogera} V.,  2019, \mn@doi [\apj]
  {10.3847/1538-4357/ab3426}, \href
  {https://ui.adsabs.harvard.edu/abs/2019ApJ...882...36M} {882, 36}

\bibitem[\protect\citeauthoryear{{McClintock}, {Narayan}  \&
  {Steiner}}{{McClintock} et~al.}{2014}]{Mc2014}
{McClintock} J.~E.,  {Narayan} R.,   {Steiner} J.~F.,  2014, \mn@doi [Space
  Science Reviews] {10.1007/s11214-013-0003-9}, \href
  {http://adsabs.harvard.edu/abs/2014SSRv..183..295M} {183, 295}

\bibitem[\protect\citeauthoryear{{Mokiem} et~al.,}{{Mokiem} et~al.}{2007}]{p46}
{Mokiem} M.~R.,  et~al., 2007, \mn@doi [A\&A] {10.1051/0004-6361:20077545},
  \href {http://adsabs.harvard.edu/abs/2007A%26A...473..603M} {473, 603}

\bibitem[\protect\citeauthoryear{{Narayan}, {Piran}  \& {Shemi}}{{Narayan}
  et~al.}{1991}]{p9}
{Narayan} R.,  {Piran} T.,   {Shemi} A.,  1991, \mn@doi [ApJL]
  {10.1086/186143}, \href {http://adsabs.harvard.edu/abs/1991ApJ...379L..17N}
  {379, L17}

\bibitem[\protect\citeauthoryear{{Neijssel} et~al.,}{{Neijssel}
  et~al.}{2019}]{Neijssel2019}
{Neijssel} C.~J.,  et~al., 2019, \mn@doi [\mnras] {10.1093/mnras/stz2840},
  \href {https://ui.adsabs.harvard.edu/abs/2019MNRAS.490.3740N} {490, 3740}

\bibitem[\protect\citeauthoryear{{Nelemans}, {Tauris}  \& {van den
  Heuvel}}{{Nelemans} et~al.}{1999}]{Nelemans1999}
{Nelemans} G.,  {Tauris} T.~M.,   {van den Heuvel} E.~P.~J.,  1999, A\&A, \href
  {http://adsabs.harvard.edu/abs/1999A%26A...352L..87N} {352, L87}

\bibitem[\protect\citeauthoryear{{Nicholl} et~al.,}{{Nicholl}
  et~al.}{2013}]{nicholl2013}
{Nicholl} M.,  et~al., 2013, \mn@doi [\nat] {10.1038/nature12569}, \href
  {https://ui.adsabs.harvard.edu/abs/2013Natur.502..346N} {502, 346}

\bibitem[\protect\citeauthoryear{{Nitz} et~al.,}{{Nitz}
  et~al.}{2019}]{Nitz:2019}
{Nitz} A.~H.,  et~al., 2019, arXiv e-prints, \href
  {https://ui.adsabs.harvard.edu/abs/2019arXiv191005331N} {p. arXiv:1910.05331}

\bibitem[\protect\citeauthoryear{{Nomoto}, {Kobayashi}  \& {Tominaga}}{{Nomoto}
  et~al.}{2013}]{Nomoto2013}
{Nomoto} K.,  {Kobayashi} C.,   {Tominaga} N.,  2013, \mn@doi [ARA\&A]
  {10.1146/annurev-astro-082812-140956}, \href
  {http://adsabs.harvard.edu/abs/2013ARA%26A..51..457N} {51, 457}

\bibitem[\protect\citeauthoryear{{O'Shaughnessy} \& {Kim}}{{O'Shaughnessy} \&
  {Kim}}{2010}]{p11}
{O'Shaughnessy} R.,  {Kim} C.,  2010, \mn@doi [ApJ]
  {10.1088/0004-637X/715/1/230}, \href
  {http://adsabs.harvard.edu/abs/2010ApJ...715..230O} {715, 230}

\bibitem[\protect\citeauthoryear{{Orosz} et~al.,}{{Orosz}
  et~al.}{2007}]{Orosz2007}
{Orosz} J.~A.,  et~al., 2007, \mn@doi [Nature] {10.1038/nature06218}, \href
  {http://adsabs.harvard.edu/abs/2007Natur.449..872O} {449, 872}

\bibitem[\protect\citeauthoryear{{Paxton}, {Bildsten}, {Dotter}, {Herwig},
  {Lesaffre}  \& {Timmes}}{{Paxton} et~al.}{2011}]{paxton2011}
{Paxton} B.,  {Bildsten} L.,  {Dotter} A.,  {Herwig} F.,  {Lesaffre} P.,
  {Timmes} F.,  2011, \mn@doi [ApJS] {10.1088/0067-0049/192/1/3}, \href
  {http://adsabs.harvard.edu/abs/2011ApJS..192....3P} {192, 3}

\bibitem[\protect\citeauthoryear{{Paxton} et~al.,}{{Paxton}
  et~al.}{2013}]{paxton2013}
{Paxton} B.,  et~al., 2013, \mn@doi [ApJS] {10.1088/0067-0049/208/1/4}, \href
  {http://adsabs.harvard.edu/abs/2013ApJS..208....4P} {208, 4}

\bibitem[\protect\citeauthoryear{{Paxton} et~al.,}{{Paxton}
  et~al.}{2015}]{paxton2015}
{Paxton} B.,  et~al., 2015, \mn@doi [ApJS] {10.1088/0067-0049/220/1/15}, \href
  {http://adsabs.harvard.edu/abs/2015ApJS..220...15P} {220, 15}

\bibitem[\protect\citeauthoryear{{Paxton} et~al.,}{{Paxton}
  et~al.}{2018}]{paxton2018}
{Paxton} B.,  et~al., 2018, \mn@doi [\apjs] {10.3847/1538-4365/aaa5a8}, \href
  {https://ui.adsabs.harvard.edu/abs/2018ApJS..234...34P} {234, 34}

\bibitem[\protect\citeauthoryear{{Paxton} et~al.,}{{Paxton}
  et~al.}{2019}]{paxton2019}
{Paxton} B.,  et~al., 2019, \mn@doi [\apjs] {10.3847/1538-4365/ab2241}, \href
  {https://ui.adsabs.harvard.edu/abs/2019ApJS..243...10P} {243, 10}

\bibitem[\protect\citeauthoryear{{Peters}}{{Peters}}{1964}]{Peters1964}
{Peters} P.~C.,  1964, \mn@doi [Physical Review] {10.1103/PhysRev.136.B1224},
  \href {http://adsabs.harvard.edu/abs/1964PhRv..136.1224P} {136, 1224}

\bibitem[\protect\citeauthoryear{{Phinney}}{{Phinney}}{1991}]{p8}
{Phinney} E.~S.,  1991, \mn@doi [ApJL] {10.1086/186163}, \href
  {http://adsabs.harvard.edu/abs/1991ApJ...380L..17P} {380, L17}

\bibitem[\protect\citeauthoryear{{Portegies Zwart} \& {McMillan}}{{Portegies
  Zwart} \& {McMillan}}{2000}]{p30}
{Portegies Zwart} S.~F.,  {McMillan} S.~L.~W.,  2000, \mn@doi [ApJL]
  {10.1086/312422}, \href {http://adsabs.harvard.edu/abs/2000ApJ...528L..17P}
  {528, L17}

\bibitem[\protect\citeauthoryear{{Prestwich} et~al.,}{{Prestwich}
  et~al.}{2007}]{Prest2007}
{Prestwich} A.~H.,  et~al., 2007, \mn@doi [ApJL] {10.1086/523755}, \href
  {http://adsabs.harvard.edu/abs/2007ApJ...669L..21P} {669, L21}

\bibitem[\protect\citeauthoryear{{Rakavy} \& {Shaviv}}{{Rakavy} \&
  {Shaviv}}{1967}]{RakaviShaviv1967}
{Rakavy} G.,  {Shaviv} G.,  1967, \mn@doi [\apj] {10.1086/149204}, \href
  {http://adsabs.harvard.edu/abs/1967ApJ...148..803R} {148, 803}

\bibitem[\protect\citeauthoryear{{Repetto}, {Igoshev}  \& {Nelemans}}{{Repetto}
  et~al.}{2017}]{repetto1}
{Repetto} S.,  {Igoshev} A.~P.,   {Nelemans} G.,  2017, \mn@doi [MNRAS]
  {10.1093/mnras/stx027}, \href
  {http://adsabs.harvard.edu/abs/2017MNRAS.467..298R} {467, 298}

\bibitem[\protect\citeauthoryear{{Rodriguez}, {Morscher}, {Pattabiraman},
  {Chatterjee}, {Haster}  \& {Rasio}}{{Rodriguez} et~al.}{2015}]{p32}
{Rodriguez} C.~L.,  {Morscher} M.,  {Pattabiraman} B.,  {Chatterjee} S.,
  {Haster} C.-J.,   {Rasio} F.~A.,  2015, \mn@doi [Phys. Rev. Letters]
  {10.1103/PhysRevLett.115.051101}, \href
  {http://adsabs.harvard.edu/abs/2015PhRvL.115e1101R} {115, 051101}

\bibitem[\protect\citeauthoryear{{Salpeter}}{{Salpeter}}{1955}]{salpeter1955}
{Salpeter} E.~E.,  1955, \mn@doi [\apj] {10.1086/145971}, \href
  {http://adsabs.harvard.edu/abs/1955ApJ...121..161S} {121, 161}

\bibitem[\protect\citeauthoryear{{Sana} et~al.,}{{Sana}
  et~al.}{2012}]{Sana2012}
{Sana} H.,  et~al., 2012, \mn@doi [Science] {10.1126/science.1223344}, \href
  {https://ui.adsabs.harvard.edu/abs/2012Sci...337..444S} {337, 444}

\bibitem[\protect\citeauthoryear{{Sathyaprakash} et~al.,}{{Sathyaprakash}
  et~al.}{2012}]{ET}
{Sathyaprakash} B.,  et~al., 2012, \mn@doi [Classical and Quantum Gravity]
  {10.1088/0264-9381/29/12/124013}, \href
  {http://adsabs.harvard.edu/abs/2012CQGra..29l4013S} {29, 124013}

\bibitem[\protect\citeauthoryear{{Sigurdsson} \& {Hernquist}}{{Sigurdsson} \&
  {Hernquist}}{1993}]{p29}
{Sigurdsson} S.,  {Hernquist} L.,  1993, \mn@doi [Nat] {10.1038/364423a0},
  \href {http://adsabs.harvard.edu/abs/1993Natur.364..423S} {364, 423}

\bibitem[\protect\citeauthoryear{{Smarr} \& {Blandford}}{{Smarr} \&
  {Blandford}}{1976}]{smarr1976}
{Smarr} L.~L.,  {Blandford} R.,  1976, \mn@doi [\apj] {10.1086/154524}, \href
  {https://ui.adsabs.harvard.edu/abs/1976ApJ...207..574S} {207, 574}

\bibitem[\protect\citeauthoryear{{Song}, {Meynet}, {Maeder}, {Ekstr{\"o}m}  \&
  {Eggenberger}}{{Song} et~al.}{2016}]{p45}
{Song} H.~F.,  {Meynet} G.,  {Maeder} A.,  {Ekstr{\"o}m} S.,   {Eggenberger}
  P.,  2016, \mn@doi [A\&A] {10.1051/0004-6361/201526074}, \href
  {http://adsabs.harvard.edu/abs/2016A%26A...585A.120S} {585, A120}

\bibitem[\protect\citeauthoryear{{Stevenson}, {Ohme}  \&
  {Fairhurst}}{{Stevenson} et~al.}{2015}]{p19}
{Stevenson} S.,  {Ohme} F.,   {Fairhurst} S.,  2015, \mn@doi [ApJ]
  {10.1088/0004-637X/810/1/58}, \href
  {http://adsabs.harvard.edu/abs/2015ApJ...810...58S} {810, 58}

\bibitem[\protect\citeauthoryear{{Stevenson}, {Sampson}, {Powell},
  {Vigna-G{\'o}mez}, {Neijssel}, {Sz{\'e}csi}  \& {Mandel}}{{Stevenson}
  et~al.}{2019}]{Stevenson2019}
{Stevenson} S.,  {Sampson} M.,  {Powell} J.,  {Vigna-G{\'o}mez} A.,  {Neijssel}
  C.~J.,  {Sz{\'e}csi} D.,   {Mandel} I.,  2019, \mn@doi [\apj]
  {10.3847/1538-4357/ab3981}, 882, 121

\bibitem[\protect\citeauthoryear{{Sz{\'e}csi}, {Langer}, {Yoon}, {Sanyal}, {de
  Mink}, {Evans}  \& {Dermine}}{{Sz{\'e}csi} et~al.}{2015}]{p44}
{Sz{\'e}csi} D.,  {Langer} N.,  {Yoon} S.-C.,  {Sanyal} D.,  {de Mink} S.,
  {Evans} C.~J.,   {Dermine} T.,  2015, \mn@doi [A\&A]
  {10.1051/0004-6361/201526617}, \href
  {http://ukads.nottingham.ac.uk/abs/2015A%26A...581A..15S} {581, A15}

\bibitem[\protect\citeauthoryear{{Taylor} \& {Kobayashi}}{{Taylor} \&
  {Kobayashi}}{2014}]{2014MNRAS.442.2751T}
{Taylor} P.,  {Kobayashi} C.,  2014, \mn@doi [MNRAS] {10.1093/mnras/stu983},
  \href {http://adsabs.harvard.edu/abs/2014MNRAS.442.2751T} {442, 2751}

\bibitem[\protect\citeauthoryear{{Taylor} \& {Kobayashi}}{{Taylor} \&
  {Kobayashi}}{2015a}]{2015MNRAS.448.1835T}
{Taylor} P.,  {Kobayashi} C.,  2015a, \mn@doi [MNRAS] {10.1093/mnras/stv139},
  \href {http://adsabs.harvard.edu/abs/2015MNRAS.448.1835T} {448, 1835}

\bibitem[\protect\citeauthoryear{{Taylor} \& {Kobayashi}}{{Taylor} \&
  {Kobayashi}}{2015b}]{2015MNRAS.452L..59T}
{Taylor} P.,  {Kobayashi} C.,  2015b, \mn@doi [MNRAS] {10.1093/mnrasl/slv087},
  \href {http://adsabs.harvard.edu/abs/2015MNRAS.452L..59T} {452, L59}

\bibitem[\protect\citeauthoryear{{Taylor} \& {Kobayashi}}{{Taylor} \&
  {Kobayashi}}{2016}]{taylor2016}
{Taylor} P.,  {Kobayashi} C.,  2016, \mn@doi [MNRAS] {10.1093/mnras/stw2157},
  \href {http://adsabs.harvard.edu/abs/2016MNRAS.463.2465T} {463, 2465}

\bibitem[\protect\citeauthoryear{{Taylor} \& {Kobayashi}}{{Taylor} \&
  {Kobayashi}}{2017}]{2017MNRAS.471.3856T}
{Taylor} P.,  {Kobayashi} C.,  2017, \mn@doi [MNRAS] {10.1093/mnras/stx1860},
  \href {http://adsabs.harvard.edu/abs/2017MNRAS.471.3856T} {471, 3856}

\bibitem[\protect\citeauthoryear{{Terreran} et~al.,}{{Terreran}
  et~al.}{2017}]{cand1}
{Terreran} G.,  et~al., 2017, \mn@doi [Nature Astronomy]
  {10.1038/s41550-017-0228-8}, \href
  {https://ui.adsabs.harvard.edu/abs/2017NatAs...1..713T} {1, 713}

\bibitem[\protect\citeauthoryear{{Toffano}, {Mapelli}, {Giacobbo}, {Artale}  \&
  {Ghirlanda}}{{Toffano} et~al.}{2019}]{toffano2019}
{Toffano} M.,  {Mapelli} M.,  {Giacobbo} N.,  {Artale} M.~C.,   {Ghirlanda} G.,
   2019, \mn@doi [\mnras] {10.1093/mnras/stz2415}, \href
  {https://ui.adsabs.harvard.edu/abs/2019MNRAS.489.4622T} {489, 4622}

\bibitem[\protect\citeauthoryear{{Tutukov} \& {Yungelson}}{{Tutukov} \&
  {Yungelson}}{1993}]{p25}
{Tutukov} A.~V.,  {Yungelson} L.~R.,  1993, \mn@doi [MNRAS]
  {10.1093/mnras/260.3.675}, \href
  {http://adsabs.harvard.edu/abs/1993MNRAS.260..675T} {260, 675}

\bibitem[\protect\citeauthoryear{{Venumadhav}, {Zackay}, {Roulet}, {Dai}  \&
  {Zaldarriaga}}{{Venumadhav} et~al.}{2019}]{iau2019}
{Venumadhav} T.,  {Zackay} B.,  {Roulet} J.,  {Dai} L.,   {Zaldarriaga} M.,
  2019, arXiv e-prints, \href
  {https://ui.adsabs.harvard.edu/abs/2019arXiv190407214V} {p. arXiv:1904.07214}

\bibitem[\protect\citeauthoryear{{Vigna-G{\'o}mez}, {Justham}, {Mandel}, {de
  Mink}  \& {Podsiadlowski}}{{Vigna-G{\'o}mez} et~al.}{2019}]{vigna2019}
{Vigna-G{\'o}mez} A.,  {Justham} S.,  {Mandel} I.,  {de Mink} S.~E.,
  {Podsiadlowski} P.,  2019, \mn@doi [\apjl] {10.3847/2041-8213/ab1bdf}, \href
  {https://ui.adsabs.harvard.edu/abs/2019ApJ...876L..29V} {876, L29}

\bibitem[\protect\citeauthoryear{{Vink} \& {de Koter}}{{Vink} \& {de
  Koter}}{2005}]{p47}
{Vink} J.~S.,  {de Koter} A.,  2005, \mn@doi [A\&A]
  {10.1051/0004-6361:20052862}, \href
  {http://adsabs.harvard.edu/abs/2005A%26A...442..587V} {442, 587}

\bibitem[\protect\citeauthoryear{{Vink}, {de Koter}  \& {Lamers}}{{Vink}
  et~al.}{2001}]{Vink2001}
{Vink} J.~S.,  {de Koter} A.,   {Lamers} H.~J.~G.~L.~M.,  2001, \aap, 369, 574

\bibitem[\protect\citeauthoryear{{Voss} \& {Tauris}}{{Voss} \&
  {Tauris}}{2003}]{p14}
{Voss} R.,  {Tauris} T.~M.,  2003, \mn@doi [MNRAS]
  {10.1046/j.1365-8711.2003.06616.x}, \href
  {http://adsabs.harvard.edu/abs/2003MNRAS.342.1169V} {342, 1169}

\bibitem[\protect\citeauthoryear{{Weisberg}, {Nice}  \& {Taylor}}{{Weisberg}
  et~al.}{2010}]{p6}
{Weisberg} J.~M.,  {Nice} D.~J.,   {Taylor} J.~H.,  2010, \mn@doi [ApJ]
  {10.1088/0004-637X/722/2/1030}, \href
  {http://adsabs.harvard.edu/abs/2010ApJ...722.1030W} {722, 1030}

\bibitem[\protect\citeauthoryear{{Woosley}}{{Woosley}}{2017}]{weasle2017}
{Woosley} S.~E.,  2017, \mn@doi [ApJ] {10.3847/1538-4357/836/2/244}, \href
  {http://adsabs.harvard.edu/abs/2017ApJ...836..244W} {836, 244}

\bibitem[\protect\citeauthoryear{{Woosley} \& {Heger}}{{Woosley} \&
  {Heger}}{2006}]{p41}
{Woosley} S.~E.,  {Heger} A.,  2006, \mn@doi [ApJ] {10.1086/498500}, \href
  {http://ukads.nottingham.ac.uk/abs/2006ApJ...637..914W} {637, 914}

\bibitem[\protect\citeauthoryear{{Woosley}, {Blinnikov}  \& {Heger}}{{Woosley}
  et~al.}{2007}]{cand2}
{Woosley} S.~E.,  {Blinnikov} S.,   {Heger} A.,  2007, \mn@doi [\nat]
  {10.1038/nature06333}, \href
  {https://ui.adsabs.harvard.edu/abs/2007Natur.450..390W} {450, 390}

\bibitem[\protect\citeauthoryear{{Yoon} \& {Langer}}{{Yoon} \&
  {Langer}}{2005}]{p40}
{Yoon} S.-C.,  {Langer} N.,  2005, \mn@doi [A\&A] {10.1051/0004-6361:20054030},
  \href {http://ukads.nottingham.ac.uk/abs/2005A%26A...443..643Y} {443, 643}

\bibitem[\protect\citeauthoryear{{Yoon}, {Langer}  \& {Norman}}{{Yoon}
  et~al.}{2006}]{Yoon2006}
{Yoon} S.-C.,  {Langer} N.,   {Norman} C.,  2006, \mn@doi [A\&A]
  {10.1051/0004-6361:20065912}, \href
  {http://adsabs.harvard.edu/abs/2006A%26A...460..199Y} {460, 199}

\bibitem[\protect\citeauthoryear{{Yusof} et~al.,}{{Yusof}
  et~al.}{2013}]{Yusof2013}
{Yusof} N.,  et~al., 2013, \mn@doi [\mnras] {10.1093/mnras/stt794}, \href
  {https://ui.adsabs.harvard.edu/abs/2013MNRAS.433.1114Y} {433, 1114}

\bibitem[\protect\citeauthoryear{{Zevin}, {Pankow}, {Rodriguez}, {Sampson},
  {Chase}, {Kalogera}  \& {Rasio}}{{Zevin} et~al.}{2017}]{zevin2017}
{Zevin} M.,  {Pankow} C.,  {Rodriguez} C.~L.,  {Sampson} L.,  {Chase} E.,
  {Kalogera} V.,   {Rasio} F.~A.,  2017, \mn@doi [\apj]
  {10.3847/1538-4357/aa8408}, \href
  {https://ui.adsabs.harvard.edu/abs/2017ApJ...846...82Z} {846, 82}

\bibitem[\protect\citeauthoryear{{de Mink} \& {Mandel}}{{de Mink} \&
  {Mandel}}{2016}]{p1}
{de Mink} S.~E.,  {Mandel} I.,  2016, \mn@doi [MNRAS] {10.1093/mnras/stw1219},
  \href {http://adsabs.harvard.edu/abs/2016MNRAS.tmp..892D} {}

\bibitem[\protect\citeauthoryear{{de Mink}, {Cantiello}, {Langer}, {Pols},
  {Brott}  \& {Yoon}}{{de Mink} et~al.}{2009}]{p34}
{de Mink} S.~E.,  {Cantiello} M.,  {Langer} N.,  {Pols} O.~R.,  {Brott} I.,
  {Yoon} S.-C.,  2009, \mn@doi [A\&A] {10.1051/0004-6361/200811439}, \href
  {http://ukads.nottingham.ac.uk/abs/2009A%26A...497..243D} {497, 243}

\makeatother
\end{thebibliography}


\appendix
\section{Grids of Detailed binary stellar evolution models}
\label{AppA}
\begin{figure*}
\includegraphics[width=\textwidth]{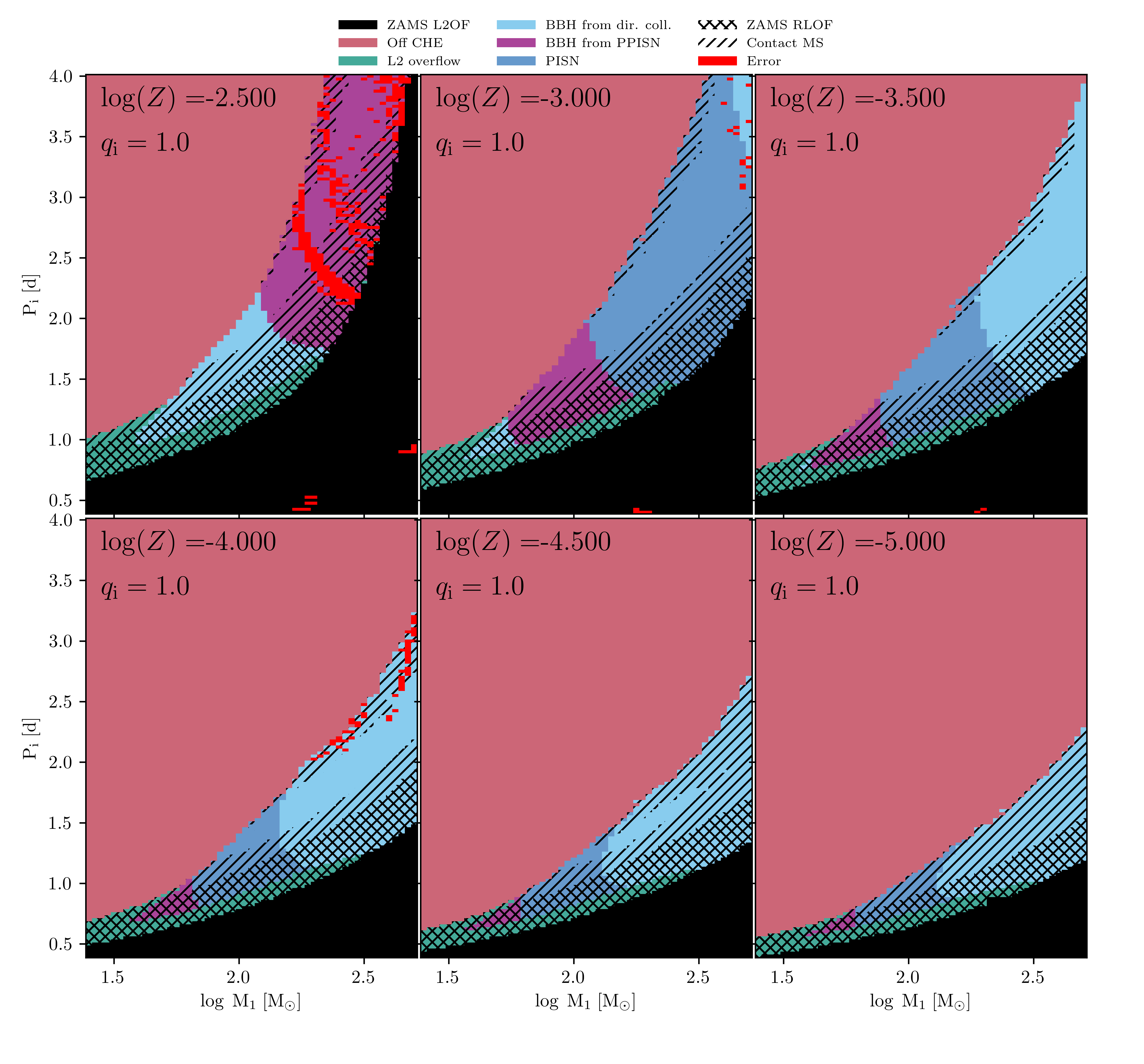}
\caption{A selection of six grids of binary systems showing their initial period and initial primary mass for $q_\textrm{i} = 1$, with final outcomes coded according to the legend at the top. Black regions show where the initial orbital period of a system was small enough to have L2 overflow at the ZAMS, while green areas are systems reaching L2 overflow during the main sequence -- systems in both these scenarios will merge during their early evolution. Pink areas depict systems that reached a point where the central and surface He abundance of one of the stars differ by more than 0.2 -- the star is therefore not evolving chemically homogeneously anymore. Blue regions successfully form double He binaries that will either collapse to form BHBH systems (light blue) or go pair-unstable (dark blue). Purple denotes areas forming BHBH systems via the PPISN path. Single hatch-marks show systems experiencing overcontact during the main sequence, while double hatch-marks are models in overcontact at the ZAMS already. Red regions show models for which the simulations didn't converge.}
\label{fig:allGrids}
\end{figure*}

\cite{p36} explored the evolution of massive close binaries by computing large grids of detailed binary evolution models using the MESA code \citep{paxton2015, paxton2013, paxton2011}, which they extended to allow for contact binaries having mass ratios close to one. Here we use similar, but more detailed grids for 22 different metallicities ranging from \mbox{$\log(Z) = -5.0$} to \mbox{$\log(Z) = -2.375$} in steps of \mbox{$\log(Z) = 0.125$}, computed for mass ratios \mbox{$q_\textrm{i} = M_2/M_1 = 1$.} These grids cover initial binary periods in the range \mbox{$0.4 \leq P_\textrm{i}/{\rm d} \leq 4.0$} in intervals of 0.025 days, and initial primary masses in the range \mbox{$1.4 \leq \log(M_1/M_\odot) \leq 2.7$} with intervals of 0.025 dex. As an example, six of these grids showing the final binary outcomes are shown in Fig. \ref{fig:allGrids}. These grids (and the rest of the data they were created from) were used to determine the final outcomes and final parameters of binary systems in our Monte Carlo simulations after their initial parameter values have been determined.

\section{Binary simulation details}
\label{AppB}
In order to perform a simulation using the cosmic star formation history of \citet{2015MNRAS.448.1835T}, we use a Monte Carlo method to sample birth redshifts and metallicities of binary systems representative of the evolution of a co-moving box. In this case, the probability of a system being formed with a redshift between $z$ and $z+dz$, and a metallicity between $\log Z$ and $\log Z+d\log Z$ is proportional to

\begin{equation}
F(z,Z)\;dz\;d\log (Z)=\frac{d^2 \mathrm{SFR}}{d \log (Z)\;dV_c} \frac{dt}{dz}\;dz\;d\log (Z),
\label{eq:sample}
\end{equation}

\noindent where $t$ represents the co-moving time coordinate and $d^2\mathrm{SFR}/d \log(Z) dV_c$ is the star formation rate per unit dex in metallicity and unit co-moving volume. To obtain a set of Monte Carlo draws satisfying this distribution function, we sample $(z_0,Z_0)$ from a flat distribution in redshift and metallicity and then reject samples with a probability equal to the ratio

\begin{equation}
p_{\textrm{reject}}(z_0,Z_0)=F(z_0,Z_0)/\textrm{Max}(F(z,Z)).
\label{eq:sample_reject}
\end{equation}

\noindent Here $\textrm{Max}$ denotes the maximum of the function $F$ in the ranges chosen for the Monte Carlo. This rejection sampling algorithm produces binaries with the desired distribution in redshift and metallicity. As the distributions for the masses, orbital periods and mass ratios of binary systems are assumed to be independent of metallicity and redshift, these are sampled separately.

\subsection{Monte Carlo Efficiency}
\label{appB_effic}
The Monte Carlo is made more efficient by sampling the binary parameters only over the ranges covered by the grids, and correcting for the uncovered parameter space. Our grids cover the following ranges:

\begin{itemize}
\item $25 \leq M_{1,\textrm{i}}/\textrm{M}_\odot \leq 500$
\item $0.4 \leq P_{\textrm{i}}/\textrm{d} \leq 4.0$
\item $q = M_2/M_1 = 1.0$
\item $-5.0 \leq \log (Z) \leq -2.375$
\end{itemize}

\noindent Integrating these distributions, we can calculate the fraction of stars in our co-moving box coming from the ranges covered by the grids. The mass, period, ratio and metallicity fractions are denoted by $C_M$, $C_P$, $C_q$ and $C_Z$, respectively. For $C_M$, assuming a Salpeter initial mass function and taking the lower mass limit for CCSNe as $8 \, \textrm{M}_{\odot}$, we find

\begin{equation}
C_M = \frac{\bigintss_{25}^{500} M^{-2.35} dM}{\bigintss_8^{\infty} M^{-2.35} dM} = 0.211,
\label{eq:Cm}
\end{equation}

\noindent and for $C_P$, taking into account that the period $P$ is drawn from a flat distribution in $\log (P)$\footnote{Note that we here assume a maximum binary period of $365.25$ days. If the reader prefers a different maximum period, $C_{P,\textrm{new}}$ can easily be recalculated and merger rates adjusted accordingly. E.g., assuming a maximum period of $4000$ days according to \cite{Sana2012}, one finds $C_{P,\textrm{new}} = 0.25$. Merger rates can then be adjusted as $R_{\textrm{new}} = R_{\textrm{old}}(C_{P,\textrm{new}}/{C_{P,\textrm{old}}})$.}, we obtain

\begin{equation}
C_P = \frac{\bigintss_{\log(0.4)}^{\log(4.0)} d\log(P)}{\bigintss_{\log(0.4)}^{\log(365.25)} d\log(P)} = 0.3378.
\label{eq:Cp}
\end{equation}

\noindent We assume that grids at $q = 1$ are representative of the range \mbox{$0.8 < q < 1.0$}, and so we have that $C_q = 0.2$. For $C_Z$, we have the effective calculation

\begin{equation}
C_Z = \cfrac{\bigint_0^{\infty} \bigint_{-5}^{-2.375} \cfrac{d^2\textrm{SFR}}{d\log (Z)\; dV_c} d\log (Z)\; dz}{\bigint_{0}^{\infty} \bigint_{-\infty}^{0} \cfrac{d^2\textrm{SFR}}{d\log (Z)\; dV_c} d\log (Z)\; dz}.
\label{eq:CZ}
\end{equation}

\noindent This integral is performed via Monte Carlo integration, and the results for $C_Z$ for our original SFR, as well as for the four modified SFR cases we consider, can be seen in Table \ref{tab:results_Cz}.

\begin{table}
\centering
\caption{The metallicity fraction $C_Z$ for the default SFR case (labeled ``Original") as well as for each of the four cases of high-redshift deviations in SFR (displayed in Figure \ref{fig:SFRdev}).}
\label{tab:results_Cz}
    \begin{tabular}{@{}lc}
    \hline
    ~        & Metallicity fraction $C_Z$    \\ \hline
    Original & $0.165$  \\
    Case 1   & $0.169$  \\
    Case 2   & $0.163$  \\
    Case 3   & $0.183$  \\
    Case 4   & $0.160$  \\ \hline
    \end{tabular}
\end{table}

\subsection{Normalization}
\label{AppB_1}
Although our Monte Carlo method provides a set of models that satisfy the assumed distribution functions, in order to compute expected rates of events it is necessary to normalize results. To do this, given a total number of simulated binaries $N_{\rm MC}$, we compute a corresponding co-moving volume $V_{\rm MC}$ for which the sampled set is representative of all binaries formed in it throughout the history of the Universe. We first define the star formation rate per unit co-moving volume as $\textrm{SFR}_V$:

\begin{equation}
\textrm{SFR}_V(z) \equiv \frac{d\textrm{SFR}}{dV_c} = \int_{-\infty}^{\infty} \frac{d^2\textrm{SFR}}{d \log (Z)\; dV_c} d\log (Z).
\label{eq:sfr_v}
\end{equation}

\noindent Following \citet{Pablo2017}, we assume the conversion constant between the mass created per unit time in stars to the co-moving supernova rate per unit co-moving volume $R_{\textrm{SN},V}(z)$ to be

\begin{equation}
\frac{R_{\textrm{SN},V} (z)}{\textrm{SFR}_V (z)} = 0.01 \textrm{M}_\odot^{-1},
\label{eq:r_sn}
\end{equation}
and that the conversion constant between $R_{\textrm{SN},V}$ and the co-moving
formation rate of massive binaries per unit co-moving volume $R_{\textrm{MB},V}$ is

\begin{equation}
R_{\textrm{MB},V} (z) = \frac{f_b}{1+f_b} R_{\textrm{SN},V} (z).
\label{eq:r_mb}
\end{equation}

\noindent Here, $f_b$ is the binary fraction, defined as the ratio between the number of binaries formed to the number of binaries and single stars formed. We assume that two out of three massive stars will form in a binary, such that $f_b = 0.5$ (\citealt{Pablo2017}). In this framework we can compute the number of massive binaries that are expected to form per unit co-moving volume during the lifetime of the Universe:

\begin{multline}
N_{\mathrm{MB},V} = \frac{f_b}{1+f_b} \left(\frac{R_{\textrm{SN},V}}{\textrm{SFR}_V}\right) \\ \int_{0}^{\infty} \int_{-\infty}^{\infty} \frac{d^2 \mathrm{SFR}}{d\log (Z)\; dV_c} \frac{dt}{dz}  d\log (Z)\; dz.
\label{eq:N_mb}
\end{multline}

\noindent Given $N_{\mathrm{MB},V}$, we can directly evaluate the value of the co-moving volume $V_{\textrm{MC}}$ representative of a number $N_{\textrm{MC}}$ of Monte Carlo draws of massive binaries,

\begin{equation}
V_{\textrm{MC}} = \frac{N_{\textrm{MC}}}{C_M C_P C_q C_Z N_{\textrm{MB},V}},
\label{eq:Vmc_last}
\end{equation}

\noindent where we have included the factors used to make the Monte Carlo sampling more efficient. Since our Monte Carlo simulation draws only massive stars, and since CCSNe have a lower mass limit of $8 \, \textrm{M}_{\odot}$, the lower mass end of the Salpeter IMF we draw from can be ignored. Considering only higher-mass stars, our Salpeter IMF is essentially equivalent to using a Kroupa IMF (similar to that used by \citealt{2015MNRAS.448.1835T}), except for a slight variation on the slope at high mass (with an exponent of 2.35 for Salpeter and 2.3 for Kroupa). Since our rates are normalised based on the number of SNe per solar mass formed, we can compare how the value of the $0.01 \, \textrm{M}_{\odot}^{-1}$ we adopted in Equation \ref{eq:r_sn} compares to a value instead derived from a Kroupa IMF. Independent of whether we consider a Kroupa IMF capped at $50 \, \textrm{M}_{\odot}$ or extrapolated to $500 \, \textrm{M}_{\odot}$, we find that the number of stars with mass $M > 8 \, \textrm{M}_{\odot}$ per solar mass of star formation is 0.010. This is in excellent agreement with our adopted value.


\subsection{Rate calculations}
\label{subsec:rates}

\noindent One can assume that for the given number of Monte Carlo draws $N_{\textrm{MC}}$, all the massive binaries ever formed in a simulation box of co-moving volume $V_{\textrm{MC}}$ have been simulated. Adding the delay times to those systems that collapse to form merging BHBHs gives the distribution of such systems in the co-moving box. Binning this distribution in redshift intervals $\Delta z$ (centred on redshifts $z_{i}$) and counting the number of BHBH mergers $N_{m,i}$ in each bin, the co-moving volumetric merger rate is

\begin{equation}
\frac{d^{2}N_m}{dtdV_c}(z_{i}) \simeq \frac{N_{m,i}}{\Delta z \times V_{\textrm{MC}}}\frac{dz}{dt}.
\label{eq:com_rate}
\end{equation}

\begin{figure}
\centering
\includegraphics[width=0.45\textwidth]{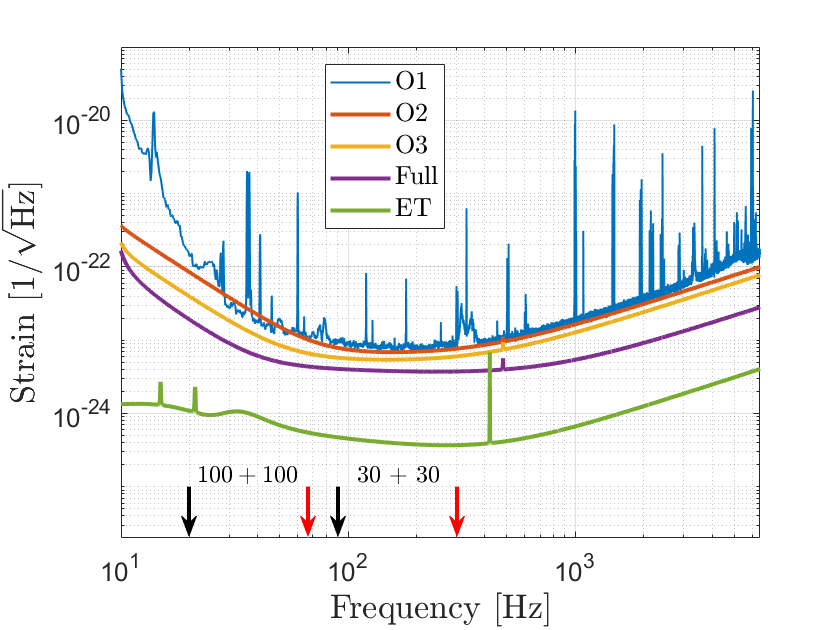}
\caption{The strain sensitivities of the O1, O2, O3 and full design sensitivity aLIGO runs \citep{aLIGOcurves2016}. Conservative predicted estimates are shown for the O2, O3 and full sensitivity runs, whereas the actual data for the O1 run are shown. The O1, O3 and full sensitivities are used when calculating the various aLIGO detection rates - the O2 sensitivity is excluded from this analysis as it is similar to that of O1. To illustrate the capabilities of future gravitational-wave detectors, we also include the Einstein Telescope's (ET) predicted sensitivity (see Section \ref{subsec:mergerrates}). The arrows at the bottom of the figure show the merger (in black) and ringdown frequency (in red, assuming a remnant spin of 0.7) of $30+30 \hspace{1mm} \textrm{M}_\odot$ and $100+100 \hspace{1mm} \textrm{M}_{\odot}$ BHBH mergers at $z=0$, respectively. The typical gravitational-wave frequency at the inspiral-merger transition is \mbox{$\sim$ 4kHz ($\textrm{M}_\odot/M$),} where $M$ is the total BHBH mass; the quasi-normal ringdown peaks at a frequency of $\sim 10-20$ kHz ($\textrm{M}_\odot/M$), depending on the final remnant spin \citep{Ajith2011}.}
\label{fig:aLIGOcurves}
\end{figure}

\noindent We are also interested in the detection rates resulting from the O1, O3 and full design aLIGO strain sensitivities, as well as the Einstein Telescope's (ET) predicted strain sensitivity. Each of these have a corresponding noise power spectrum $S_\textrm{n}$, which can be seen in Figure \ref{fig:aLIGOcurves}. Given the merging redshift and masses for a BHBH merger, we can estimate its detection probability $p_k$ for each different noise power spectrum - see Appendix \ref{appB_detector} for a description of how this is calculated. The co-moving rate of mergers per unit co-moving volume $V_{\textrm{MC}}$ detectable locally at $z=0$ is then obtained by adding the detection probabilities of all mergers in a redshift bin (denoted by $N_{det,i}$) to obtain 

\begin{equation}
\frac{d^{2}N_{det}}{dtdV_c}(z_{i}) \simeq \frac{N_{det,i}}{\Delta z \times V_{\textrm{MC}}}\frac{dz}{dt},
\label{eq:aligo_com_rate}
\end{equation}

\noindent similar to what was done before. In the case of all detection probabilities equalling $p_k = 1.0$, Equation \ref{eq:com_rate} and \ref{eq:aligo_com_rate} are equivalent. The co-moving volumetric merger rate $dN_m / dtdV_c$ can next be transformed into a cosmological rate observable from Earth as 

\begin{equation}
R_m = \int_{0}^{\infty} \frac{1}{1+z} \frac{d^{2}N_m}{dtdV_c} \frac{dV_c}{dz} dz,
\label{eq:rm_main}
\end{equation}

\noindent which in practice can be computed as

\begin{equation}
R_m \simeq \sum_{i} \frac{1}{1+z_i} \frac{d^{2}N_m}{dtdV_c} \frac{dV_c}{dz} \Delta z.
\label{eq:rm2_main}
\end{equation}

\noindent To calculate PISNe rates, the procedure is equivalent, except that the redshift of the event is the same as its formation redshift, and each of our binary models produce two PISNe.

\subsection{Testing of Rate Calculations}
\label{norm_test}

In order to verify that our setup to compute rates works as expected, we perform a synthetic test for which we know what the resulting rate should be. Given the boundaries of our grid of binary models, and the assumption that our $q=1$ models are representative of systems with $0.8<q<1$, the co-moving rate of formation of binaries covered by our grids is

\begin{eqnarray}
\begin{aligned}
R_{{\rm grid}, V} (z) = 0.01M_\odot^{-1} \times C_{M}C_{P}C_{q}\frac{f_b}{1+f_b}\qquad\qquad\qquad \\
\int_{-5}^{-2.375} \frac{d^2\textrm{SFR}}{d \log (Z)\; dV_c} d\log (Z).
\end{aligned}
\end{eqnarray}

\noindent If we were to ignore our binary evolution calculations and instead assume all binaries in the range of our grids form a merging BHBH with zero time delay between the time when the stars are formed and the moment the BHs merge, then the co-moving volumetric merger rate would satisfy

\begin{eqnarray}
\frac{d^{2}N_m}{dtdV_c}(z) = R_{{\rm grid}, V} (z).
\end{eqnarray}

\begin{figure}
\centering
\includegraphics[width=0.45\textwidth]{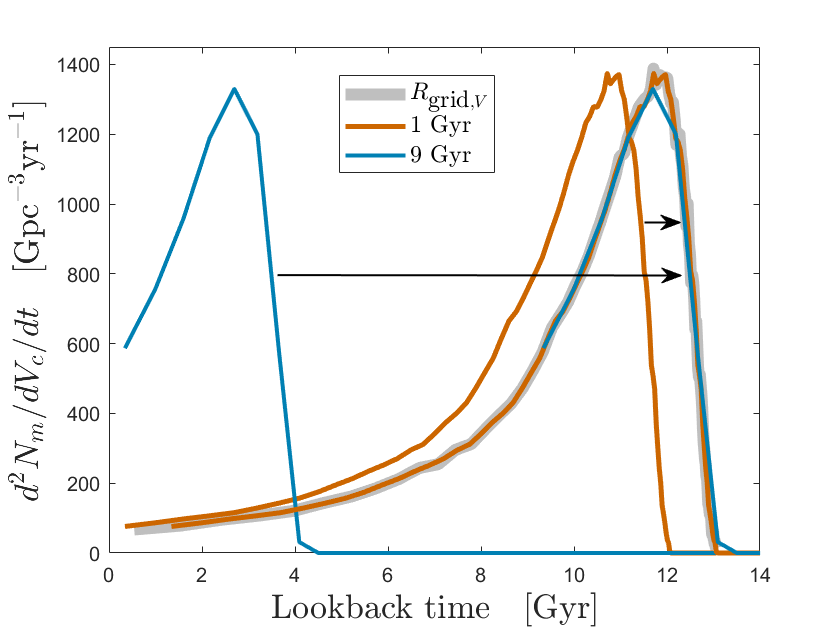}
\caption{Synthetic test performed to check our method of computing merger rates. See the text of Section \ref{norm_test} for details. The 1 Gyr and 9 Gyr test curves are also shown shifted (as indicated by the arrows) to asses how well their shapes match that of $R_{\textrm{grid}, V}$.}
\label{fig:test}
\end{figure}

Moreover, if one assumes all merging BHBHs have a fixed delay time between formation and merger, then the co-moving volumetric rate of mergers would match $R_{{\rm grid},V}(z)$ shifted in lookback time by the amount chosen for the delay time. This allows us to test our calculations, since we can directly use the star formation data to compute $R_{{\rm grid},V}(z)$ and compare to the resulting co-moving volumetric rate from our Monte Carlo calculations. The result of this exercise is displayed in Fig. \ref{fig:test}, where we show the outcome of assuming fixed delay times of $1$ Gyr and $9$ Gyrs. Shifting the predicted rates by the assumed time delay matches exactly $R_{{\rm grid},V}(z)$. In these calculations we also assumed that the mass of the BHs formed matched the initial mass of their progenitor stars, and verified that the mass distribution of merging BHBHs matched the Salpeter distribution used for the stellar masses.

\subsection{Detection Probability}
\label{appB_detector}
In order to calculate the merger rate detectable by aLIGO, the probability of detection of each one of our simulated BHBH systems that merge in the lifetime of the Universe needs to be determined. To do this, an isotropic Universe is assumed and hence each binary can be located anywhere in the sky with equal probability. Given that the sensitivity of aLIGO is dependent on the location of the binary in the sky, the signal from each simulated binary will have a different signal to noise ratio (SNR) depending on its location in the sky. The probability of detection $p_k$ for a BHBH merger is hence defined as the fractional area in the sky where its SNR achieved a sufficiently high value to yield a reliable detection, and can be written as

\begin{equation}
p_k = 1 - \mathcal{C} \left[ \min \left( \frac{8}{n_{\textrm{SNR}}}, 1 \right) \right] .
\label{eq:ilja1}
\end{equation}

\noindent Here, we have adopted a SNR detection threshold of 8, and $\mathcal{C}$ is the inverse of the cumulative distribution function of projection coefficients (and is an integral over inclination, orientation and sky location) and can be used as a representation of detector sensitivity (see \citealt{p1}). The maximum SNR $n_{\textrm{SNR}}$ for a specific BHBH merger is calculated in Fourier space and is determined by the merger's chirp mass $M_{\textrm{chirp}}$ and redshift (which determines the strength at which the signal will arrive at Earth), as well as the noise properties of the aLIGO detector. This can be expressed as

\begin{equation}
n_{\textrm{SNR}}^2 (z_m, M_{\textrm{chirp}}) = 4 \int\limits_0^\infty \frac{| h(f,z_m,M_{\textrm{chirp}}) |^2}{S_\textrm{n}(f)} df,
\label{eq:ilja2}
\end{equation}

\noindent where $h$ is the waveform of the face-on, overhead gravitational-wave signal of a BHBH merger placed at redshift $z_m$ with chirp mass $M_{\textrm{chirp}}$, and $S_\textrm{n}$ is the noise power spectrum of the detector. The chirp mass is defined as

\begin{equation}
M_{\textrm{chirp}} = \frac{(M_1 M_2)^{3/5}}{(M_1 + M_2)^{1/5}}.
\label{eq:chirpmass}
\end{equation}

\noindent The descriptions and assumptions in this section are based on the work of \cite{p1}.

\label{lastpage}
\end{document}